\documentclass[runningheads]{svmult}

\usepackage{makeidx}   % allows index generation
\usepackage{graphicx}  % standard LaTeX graphics tool
\usepackage{subeqnar}  % subnumbers individual equations
\usepackage{multicol}  % used for the two-column index
\usepackage{physprbb}  % modified textarea for proceedings,
\makeindex             % used for the subject index
\newcommand{\hgamma}{\mbox{H$\gamma$}}

\def\fre#1{\frac{ {\hbox{$ #1 $}} } }
\def\fri#1{ {\hbox{$ #1 $}}  }
\begin{document}
\title*{Blue Supergiants as a Tool for Extragalactic Distances -- 
Theoretical Concepts
\footnote{Invited review at the International Workshop on \emph{Stellar
Candles for the Extragalactic Distance Scale}, held in Concepci\'on, Chile,
December 9--11, 2002. To be published in: \emph{Stellar Candles}, Lecture
Notes in Physics (http://link.springer.de/series/lnpp), Copyright:
Springer-Verlag, Berlin--Heidelberg--New York, 2003}
}

\toctitle{
Blue Supergiants as a Tool for Extragalactic Distances}
% allows explicit linebreak for the table of content
%
%
\titlerunning{Blue Supergiants as a Tool for Extragalactic Distances}
% allows abbreviation of title, if the full title is too long
% to fit in the running head
%
\author{Rolf-Peter Kudritzki and Norbert Przybilla}
\authorrunning{R.P. Kudritzki and N. Przybilla}
% if there are more than two authors,
% please abbreviate author list for running head
%
%
\institute{Institute for Astronomy, University of Hawaii, 
Honolulu HI 96822, USA}

\maketitle              % typesets the title of the contribution

\begin{abstract}
Because of their enormous intrinsic brightness blue supergiants are ideal
stellar objects to be studied spectroscopically as individuals in galaxies far
beyond the Local Group. Quantitative spectroscopy by means of efficient
multi-object spectrographs attached to 8m-class telescopes and modern NLTE
model atmosphere techniques allow us to determine not only intrinsic stellar
parameters such as effective temperature, surface gravity, chemical 
composition and
absolute magnitude but also very accurately interstellar reddening and
extinction. This is a significant advantage compared to classical distance
indicators like Cepheids and RR\,Lyrae. We describe the spectroscopic
diagnostics of blue supergiants and introduce two concepts to determine
absolute magnitudes. The first one (Wind Momentum -- Luminosity Relationship)
uses the correlation between observed stellar wind momentum and
luminosity, whereas the second one (Flux-weighted Gravity -- Luminosity
Relationship) relies only on the determination of effective temperature and 
surface gravity to yield an accurate estimate of absolute magnitude. We 
discuss the potential of these two methods. 
\end{abstract}

%========================================================================

\section{Introduction}

The best established stellar distance indicators, Cepheids and RR\,Lyrae,
suffer from two major problems, extinction and metallicity dependence, both 
of which
are difficult to determine for these objects with sufficient precision. Thus,
in order to improve distance determinations in the local universe and to
assess the influence of systematic errors there is
definitely a need for alternative distance indicators, which are at least as
accurate but are not affected by uncertainties arising from extinction or
metallicity. It is our conviction that blue supergiants are ideal objects for 
this purpose. The big advantage is the enormous intrinsic brightness in 
visual light, which makes them available for accurate quantitative 
spectroscopic studies even far beyond the Local Group using the new generation
of 8m-class telescopes and the extremely efficient multi-object spectrographs 
attached to them \cite{bresolin01}. Quantitative spectroscopy allows us to
determine the stellar parameters and thus the intrinsic energy distribution,
which can then be used to measure reddening and the extinction law. In
addition, metallicity can be derived from the spectra. We emphasize that a 
reliable {\em spectroscopic}
distance indicator will always be superior, since an enormous amount of
additional information comes for free, as soon as one is able to obtain a
reasonable spectrum.

In this review we concentrate on blue supergiants of spectral types late B
to early A. These are the the brightest ``normal'' stars at
visual light with absolute magnitudes $-7.0  \ge M_{V} \ge -9.5$,
see \cite{bres03}. By ``normal'' we mean stars evolving peacefully without 
showing signs of eruptions or explosions, which are difficult to handle 
theoretically and observationally.

\begin{figure}[t]
\begin{center}
\includegraphics[width=.775\textwidth]{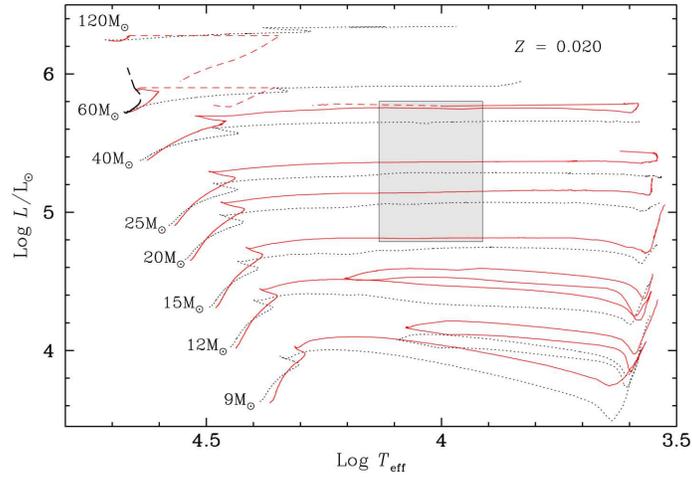}
\end{center}
\caption[]{
Evolutionary tracks of massive stars in the Hertzsprung-Russell 
diagram and the location of blue supergiants of late B and early A spectral
types ({\it shaded box}). The tracks are from \cite{meynet00}. Solid tracks 
include the effects
of stellar rotation, whereas dotted tracks are for non-rotating stars. 
Solar metallicity has been adopted for the calculations. The tracks are
labelled by the initial masses (in solar units) at the zero-age main sequence
(ZAMS). Note that effects of mass-loss due to stellar winds are included in
the calculations so that actual masses are smaller than ZAMS-masses in later
stages of the evolution
}
\label{hrd}
\end{figure}

Figure~\ref{hrd} shows the location of these objects in a Hertzsprung-Russell
diagram (HRD) with theoretical evolutionary tracks. With initial ZAMS-masses
between 15 and 40\,M$_{\odot}$ they do not belong to the most massive and the 
most luminous stars in galaxies. O-stars can be significantly more massive
and luminous, however, because of their high atmospheric temperatures they 
emit most of their radiation in the extreme and far UV. Late B and early A
supergiants are cooler and because of Wien's law their bolometric
corrections are small so that their brightness at visual light reaches a
maximum value during stellar evolution.

In the temperature range of late B and early A-supergiants there are also
always a few objects brighter than M$_{V}$\,$=$\,$-$9.5 . Generally, those are 
more exotic objects such as Luminous Blue Variables (LBVs) with higher
initial masses and with spectra characterized by strong emission lines and
sometimes in dramatic evolutionary phases with outbursts and eruptions.
Although their potential as distance indicators is also very promising, we
regard the physics of their evolution and atmospheres as too complicated at
this point and, thus, exclude them from our discussion.

The objects of our interest evolve smoothly from the left to the right in
the HRD crossing the temperature range of late B and early A-supergiants on
the order of several $10^{3}$ years \cite{meynet00}. During this short evolutionary 
phase stellar winds with mass-loss rates of the order 
$10^{-6}$ M$_{\odot}$\,yr$^{-1}$ or less
\cite{kud2000} do not have time enough to reduce the mass of the star 
significantly so that the mass remains constant. In addition, as 
Fig.~\ref{hrd} shows, the luminosity
stays constant as well. The fact that the evolution of these objects can 
very simply be described by constant mass, luminosity and a straightforward
mass-luminosity relationship makes them a very attractive stellar distance
indicator, as we will explain later in this review.

As evolved objects the blue supergiants are older than their O-star
progenitors, with ages between 0.5 to 1.3\,$\times$\,$10^{7}$ years 
\cite{meynet00}. 
All galaxies with ongoing star formation or bursts of this age will show
such a population. Because of their age they are spatially less concentrated
around their place of birth than O-stars and can frequently be found as 
isolated field stars. This together with their intrinsic brightness makes them
less vulnerable as distance indicators against the effects of crowding even at
larger distances, where less luminous objects such as Cepheids and RR\,Lyrae
start to have problems.

With regard to the crowding problem we also note that the short evolutionary
time of $10^{3}$ years makes it generally very unlikely that an unresolved
blend of two supergiants with very similar spectral types is observed. On
the other hand, since we are dealing with spectroscopic distance indicators,
any contribution of unresolved additional objects of different spectral type
is detected immediately, as soon as it affects the total magnitude 
significantly.

Thus, it is very obvious that blue supergiants seem to be ideal to investigate
the properties of young populations in galaxies. They can be used to study
reddening laws and extinction, detailed chemical composition, i.e. not only
abundance patterns but also gradients of abundance patterns as a function of
galactocentric distance, the properties of stellar winds as function of
chemical composition and the evolution of stars in different galactic
environment. Most importantly, as we will demonstrate below, they are
excellent distance indicators.

It is also very obvious that the use of blue supergiants as tools to
understand the physics of galaxies and to determine their distances depends
very strongly on the accuracy of the spectral diagnostic methods which are
applied. The attractiveness of blue supergiants for extragalactic work,
namely their outstanding intrinsic brightness, has also always posed a
tremendous theoretical problem. The enormous energy and momentum density
contained in their photospheric radiation field leads to significant
departures from Local Thermodynamic Equilibrium and to stellar wind outflows
driven by radiation. It has long been a problem to model non-LTE and
radiation driven winds realistically, but significant theoretical progress was
made during the past decade resulting in powerful spectroscopic diagnostic
tools which allow to determine the properties of supergiant stars with high
precision. 

We describe the status quo of the spectroscopic diagnostics in
the following chapters. We will then demonstrate how the spectroscopic
information can be used to determine distances. We will introduce two
completely independent theoretical concepts for distance determination methods.
The first method, the
Wind Momentum -- Luminosity Relationship (WLR), uses the strengths of the
radiation driven stellar winds as observed through the diagnostics of 
H$_{\alpha}$ as a measure of absolute luminosity and, therefore, distance. The
second method, the Flux-weighted Gravity -- Luminosity Relationship (FGLR),
determines the stellar gravities from all the higher Balmer lines and uses
gravity divided by the fourth power of effective temperature as a precise
measure of absolute magnitude. A short discussion of the potential of these new
concepts will conclude the paper.

\section{Stellar Atmospheres and Spectral Diagnostics}

The physics of the atmospheres of blue supergiant stars is
complex and very different from standard stellar atmosphere models. They   
are dominated by the influence of the radiation field, which is characterized 
by energy densities larger than or of the same order as the energy 
density of 
atmospheric matter. Another important characteristic are the low gravities,
which lead to extremely low densities and an extended atmospheric plasma
with very low escape velocity from the star. 
This has two important consequences. First, severe departures from Local 
Thermodynamic Equilibrium (LTE) of the level populations in the entire atmosphere
are induced, because radiative transitions between ionic energy levels become
much more important than inelastic collisions with free electrons. Second, 
a supersonic 
hydrodynamic outflow of atmospheric matter is initiated by line 
absorption of photons transferring outwardly directed momentum to the 
atmospheric plasma. This latter mechanism is responsible for the existence of 
the strong stellar winds observed. 

Stellar winds can affect the structure of the outer atmospheric layers
substantially and change the profiles of strong optical lines such as 
H$_{\alpha}$ and H$_{\beta}$ significantly.
The effects of the departures from Local Thermodynamic Equilibrium 
(``NLTE'') can also become crucial depending on the atomic properties of the
ion investigated. A comprehensive and detailed discussion of the basic
physics behind these effects and the advancement of model atmosphere work
for blue supergiants is given in \cite{kud88} and~\cite{kud98}. More recent work
is described in \cite{santolaya97},\cite{paul2001} and~\cite{puls2002}.

\begin{figure}[!ht]
\begin{center}
\resizebox{.45\hsize}{!}{\includegraphics{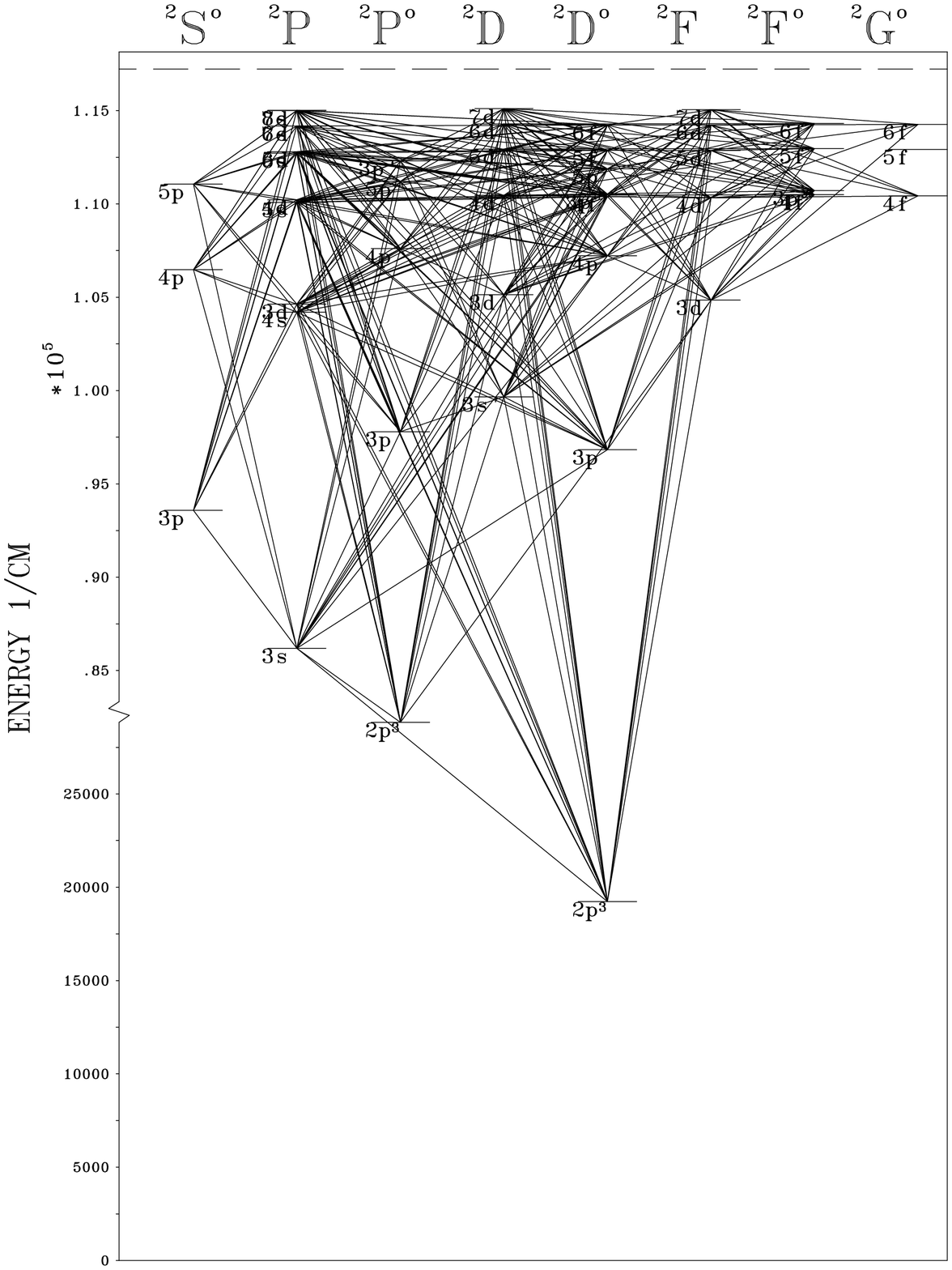}}
\hfill
\resizebox{.45\hsize}{!}{\includegraphics{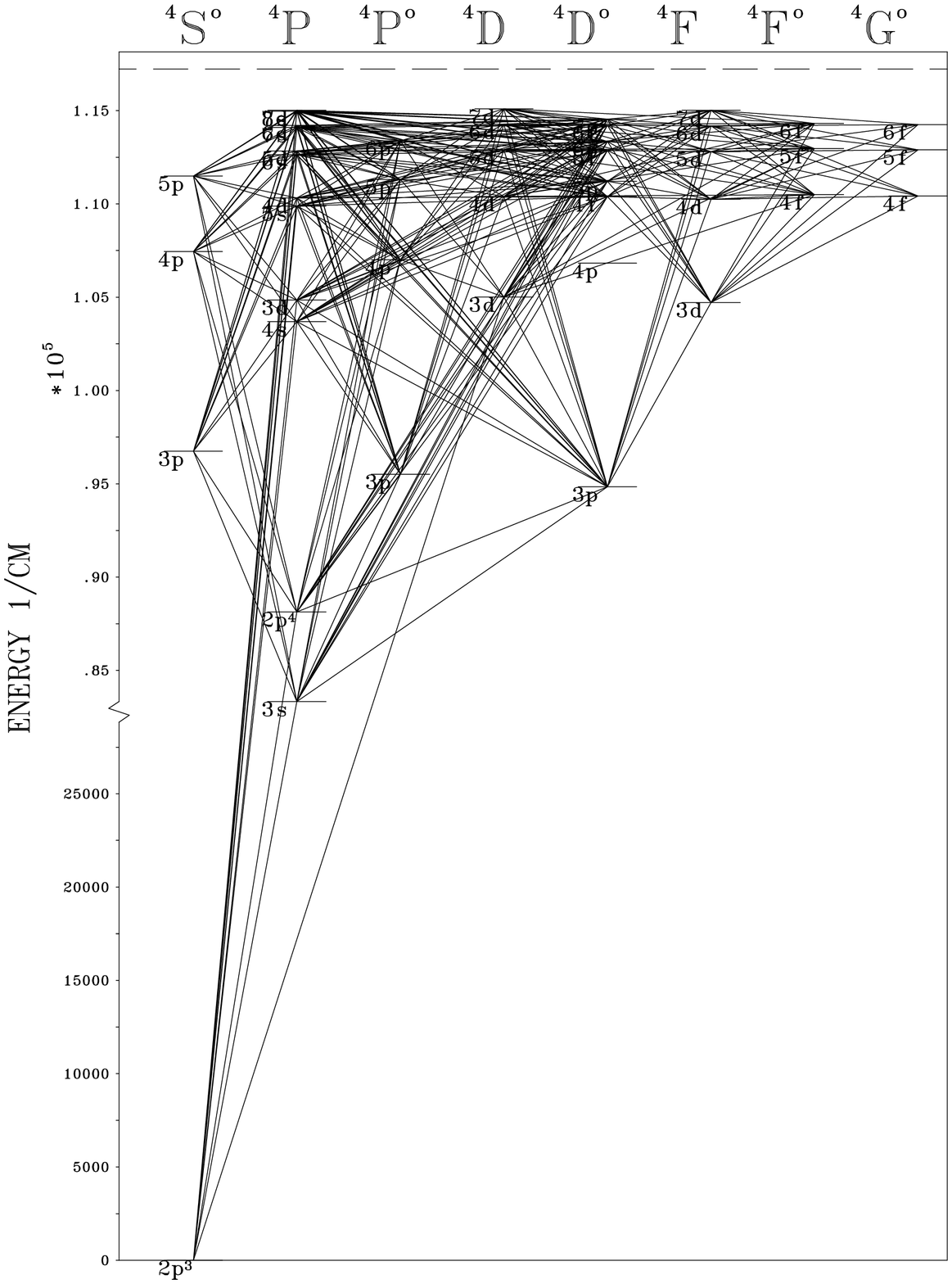}}\\[-.9cm]
\resizebox{.45\hsize}{!}{\includegraphics{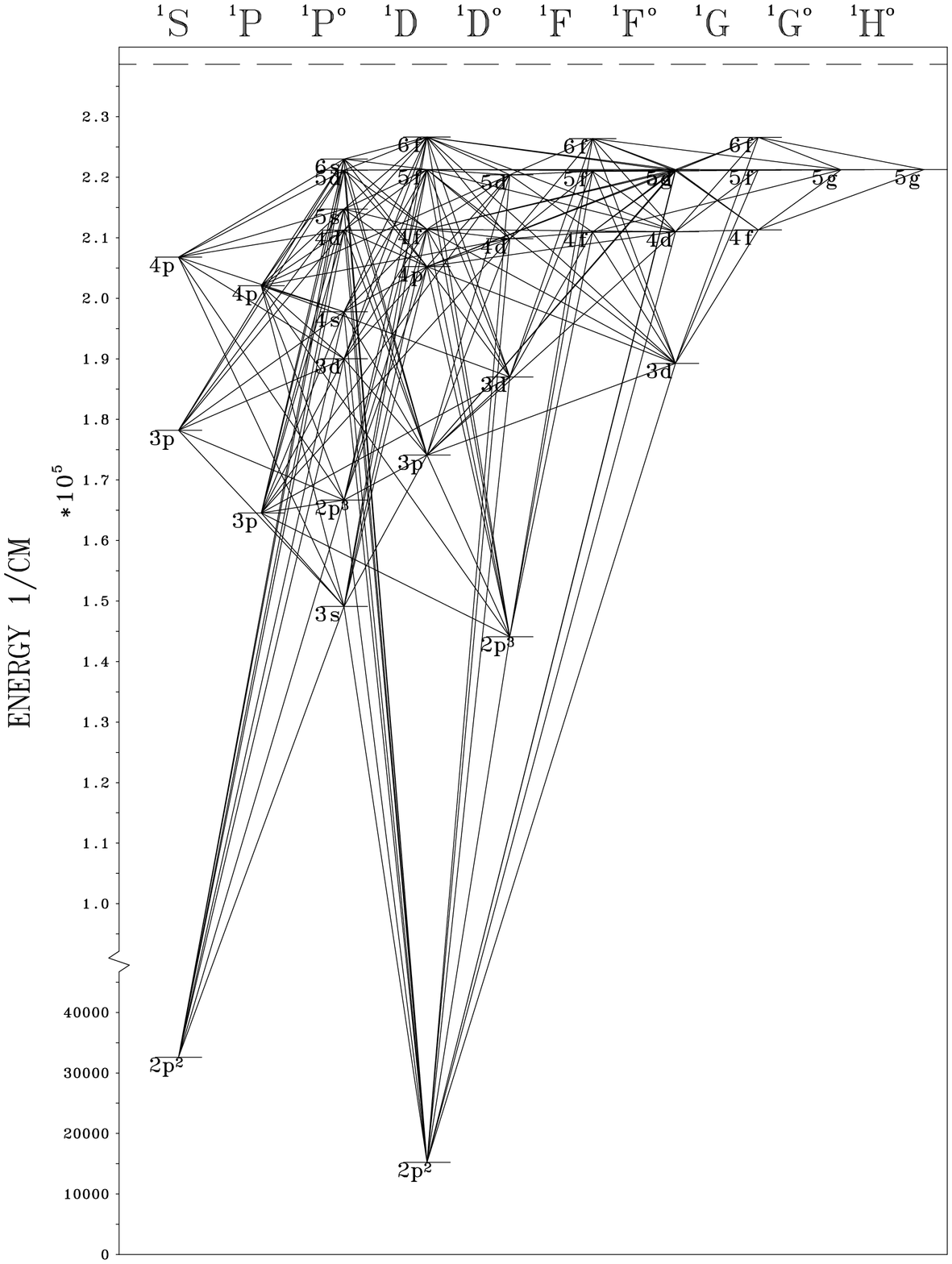}}
\hfill
\resizebox{.45\hsize}{!}{\includegraphics{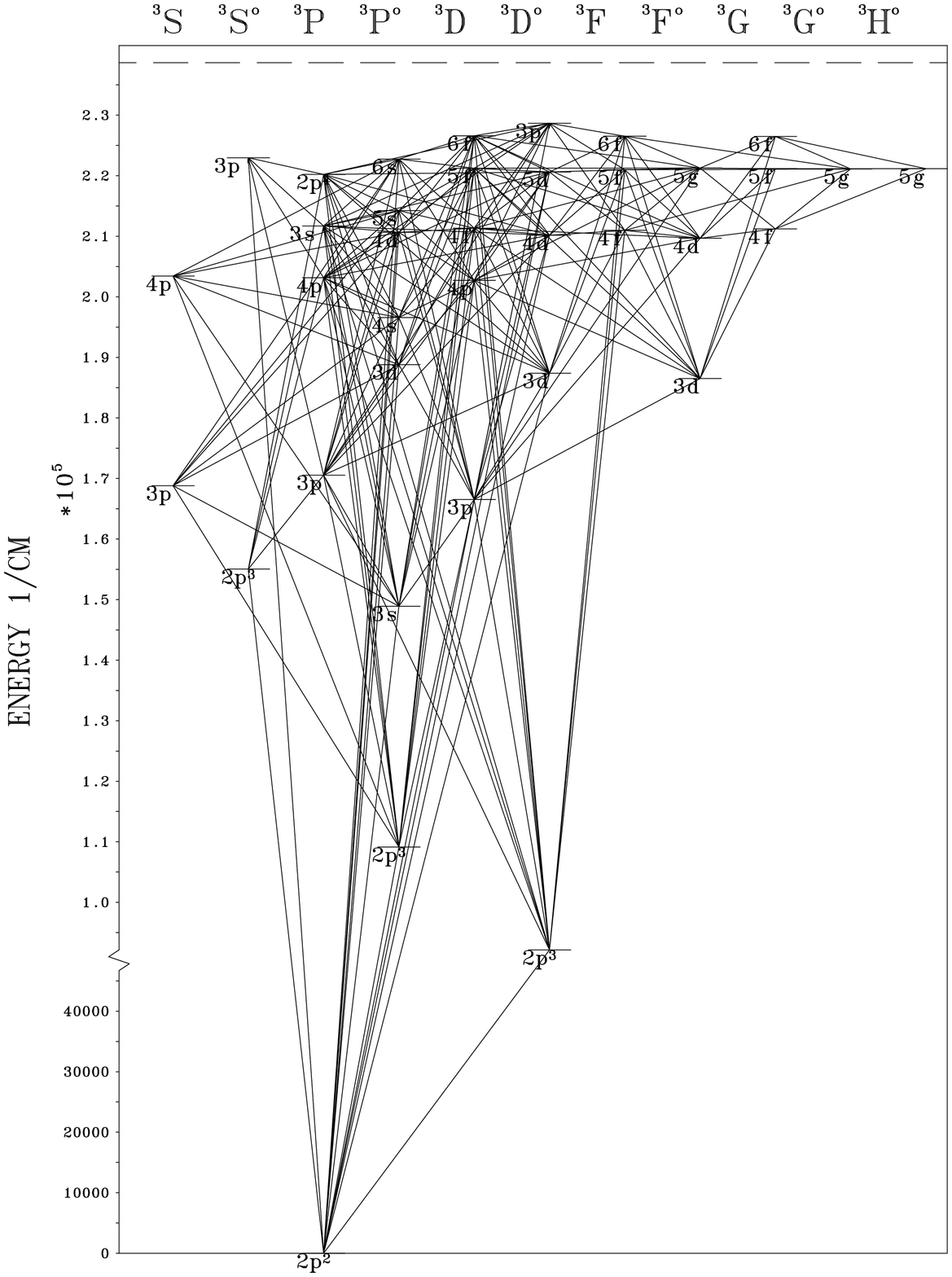}}%\\[-8.5mm]
\end{center}
\caption[]{
Model atoms for N\,{\sc i} (top) and N\,{\sc ii} (bottom) as used in the NLTE model
calculations for late B and early A-supergiants; from \cite{przy01a}
}
\label{nitro}
\end{figure}

For late B and early A-supergiants considerable progress has been made
during the last four years in
the development of a very detailed and accurate modelling of the NLTE
radiative transfer enabling very precise determinations of stellar parameters
and chemical abundances, see \cite{becker98}, \cite{przy00}, \cite{przy01a},
\cite{przy01b}, \cite{przy01c} and~\cite{przy02}. Figure~\ref{nitro} gives an
impression of the effort put into the atomic models and corresponding
radiative transfer of individual ions. 

\subsection{Effective Temperature and Gravity}

\begin{figure}[t]
\begin{center}
\includegraphics[width=.7\textwidth]{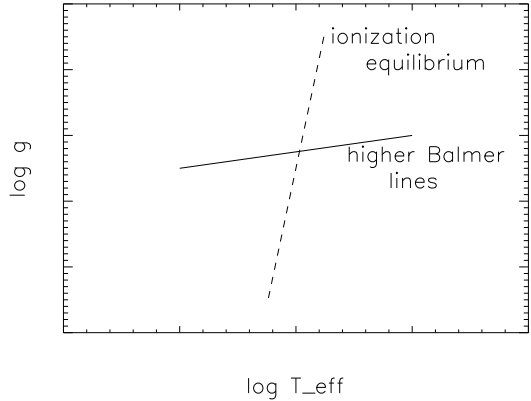}
\end{center}
\caption[]{
Schematic fit diagram of temperature- and gravity-sensitive spectral lines
in the ($\log g$, log T$_{\mathrm{eff}}$)-plane. Along the dashed curve the computed
spectral lines of two different ionization stages of one element agree with
the observations. Typical ionization equilibria for late B-supergiants are
Si\,{\sc ii/iii}, N\,{\sc i/ii}, O\,{\sc i/ii} and S\,{\sc ii/iii},
and for A-supergiants one can use Mg\,{\sc i/ii} and N\,{\sc i/ii}. 
Along the solid curve the computed profiles of the higher Balmer
lines agree with the observations. The intersection determines 
T$_\mathrm{eff}$ and $\log g$
}
\label{ftg}
\end{figure}

Effective temperature T$_\mathrm{eff}$ and gravity $\log g$ are the most fundamental
atmospheric parameters.
They are usually determined by fitting simultaneously two sets of spectral
lines, one depending mostly on T$_\mathrm{eff}$ and the other on $\log g$.
Figure~\ref{ftg} indicates how this is done in principle.
When fitting the ionization equilibria of elements spectral lines of two or
more
ionization stages have to be brought into simultaneous agreement with
observations. At different locations in the
($\log g$, log T$_\mathrm{eff}$)-plane this can be achieved only for 
different elemental
abundances. Thus, along the fit curve for the ionization equilibrium in 
Fig.~\ref{ftg}
the abundance of the corresponding element varies and the intersection with
the fit curve for the Balmer lines leads to an automatic determination of
the abundance of the element, for which the ionization equilibrium is
investigated. (Note that the
old technique of fitting ratios of equivalent widths of lines in different
ionization stages and to regard those as being independent of abundance is less
reliable, since the lines might be on different parts on the curve of growth).

\begin{figure}[t]
\begin{center}
\includegraphics[width=.65\textwidth]{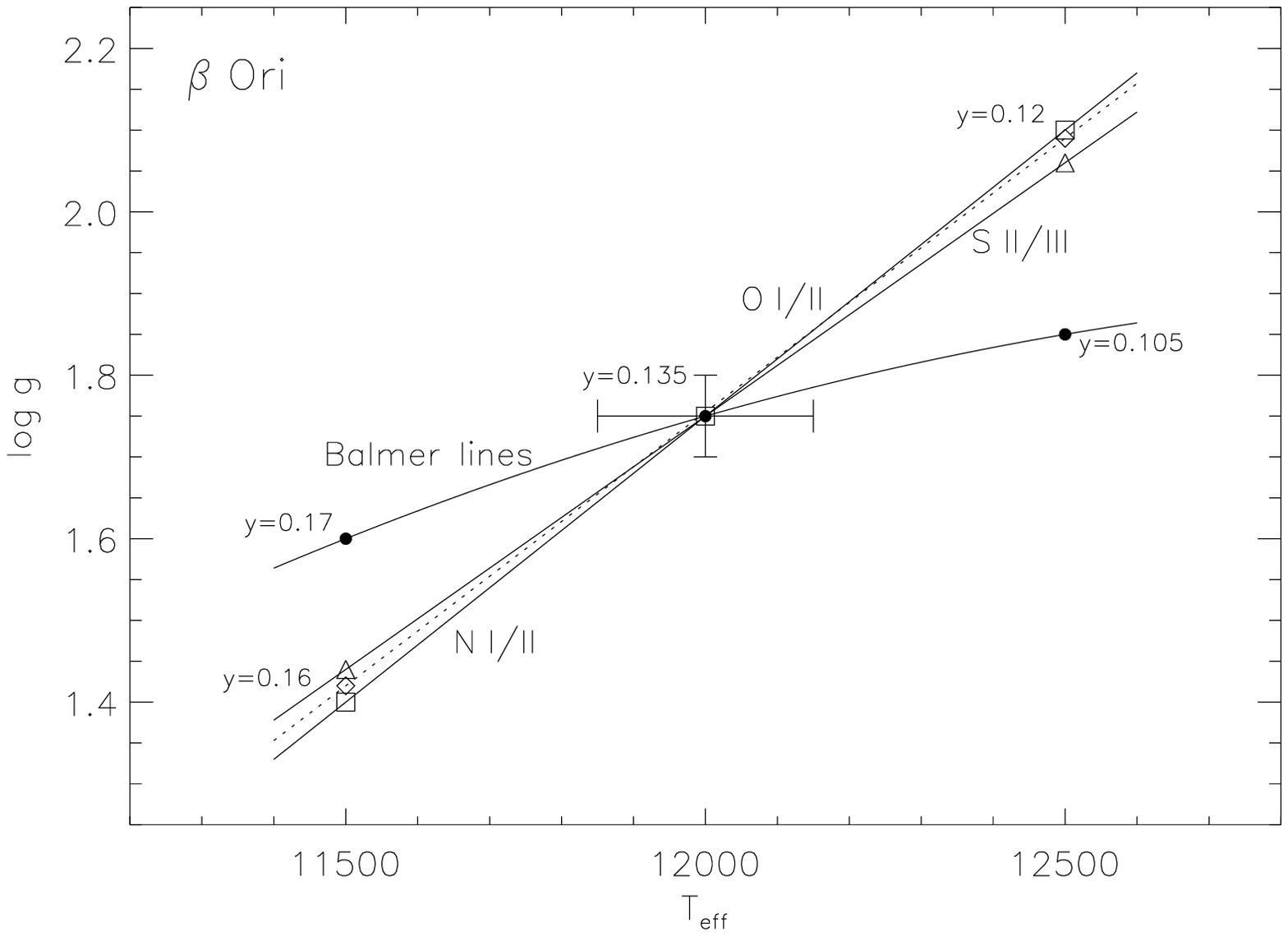}\\[-4mm]
\includegraphics[width=.65\textwidth]{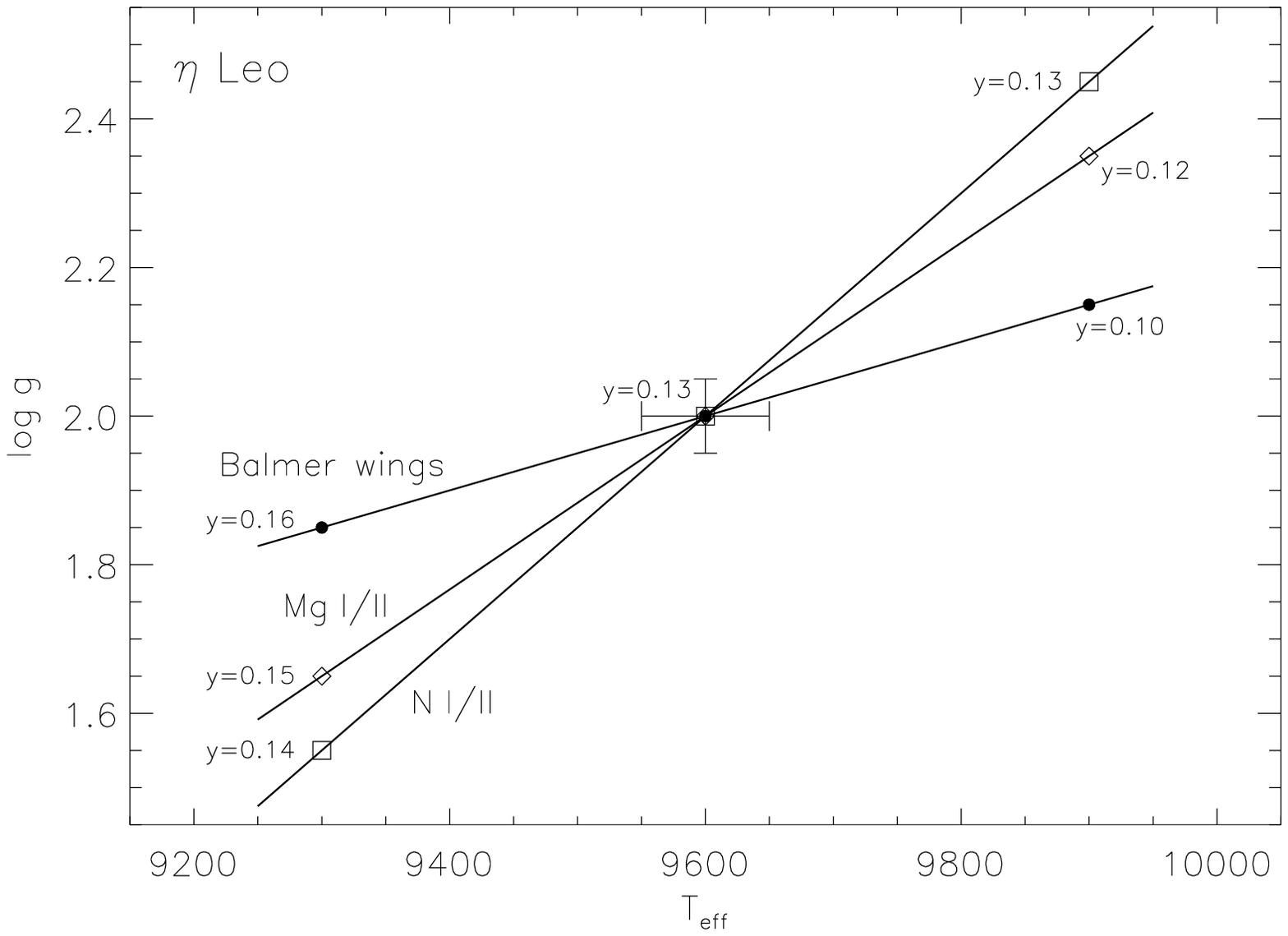}
\end{center}
\caption[]{
Fit diagram for the supergiants $\beta$\,Ori (top) and $\eta$\,Leo (bottom).
The curves are parameterized by surface helium abundance $y$ (by number);
from \cite{przy02}
}
\label{ftg1}
\end{figure}

\begin{figure}[ht]
\begin{center}
\includegraphics[width=.7\textwidth]{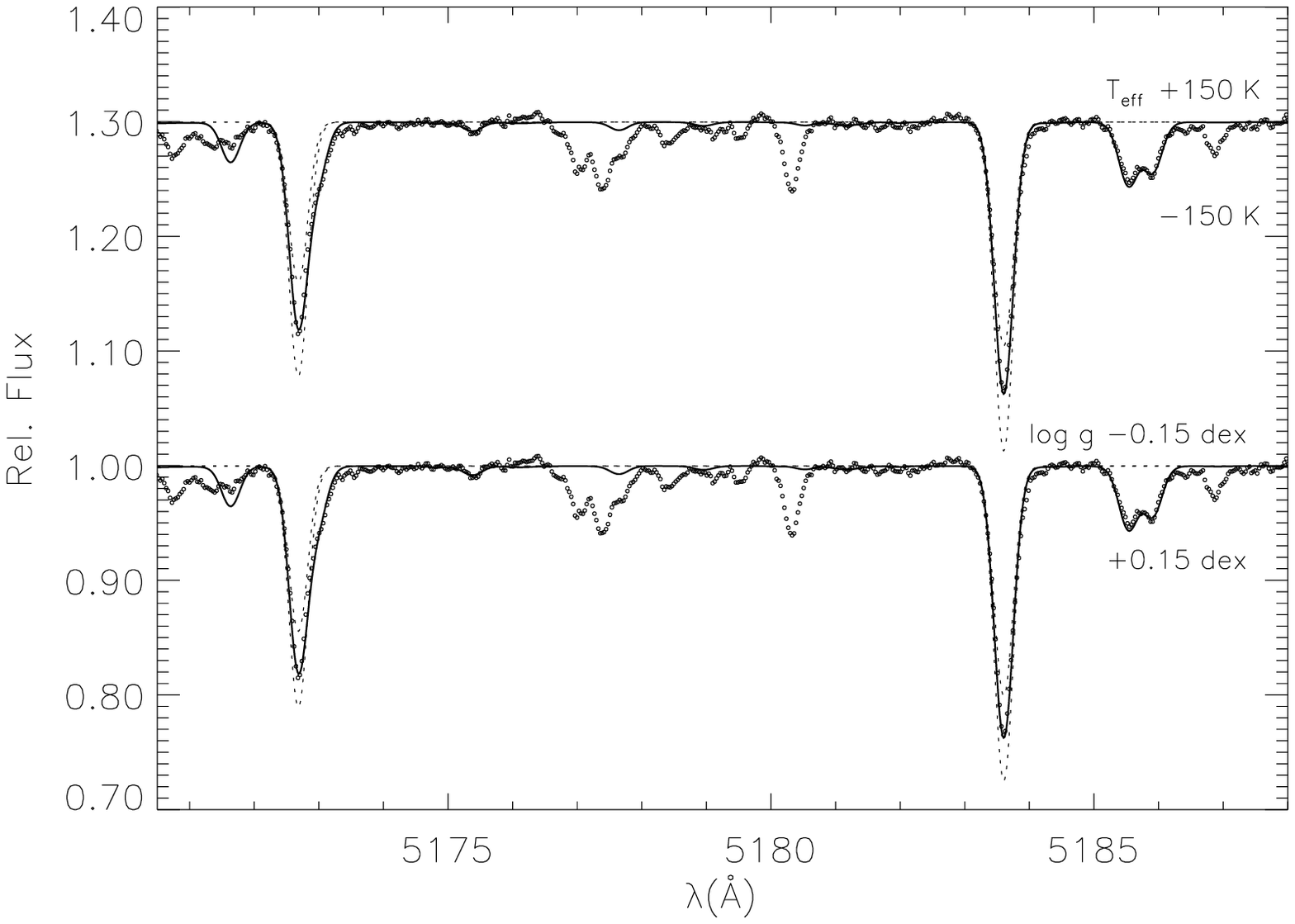}
\includegraphics[width=.7\textwidth]{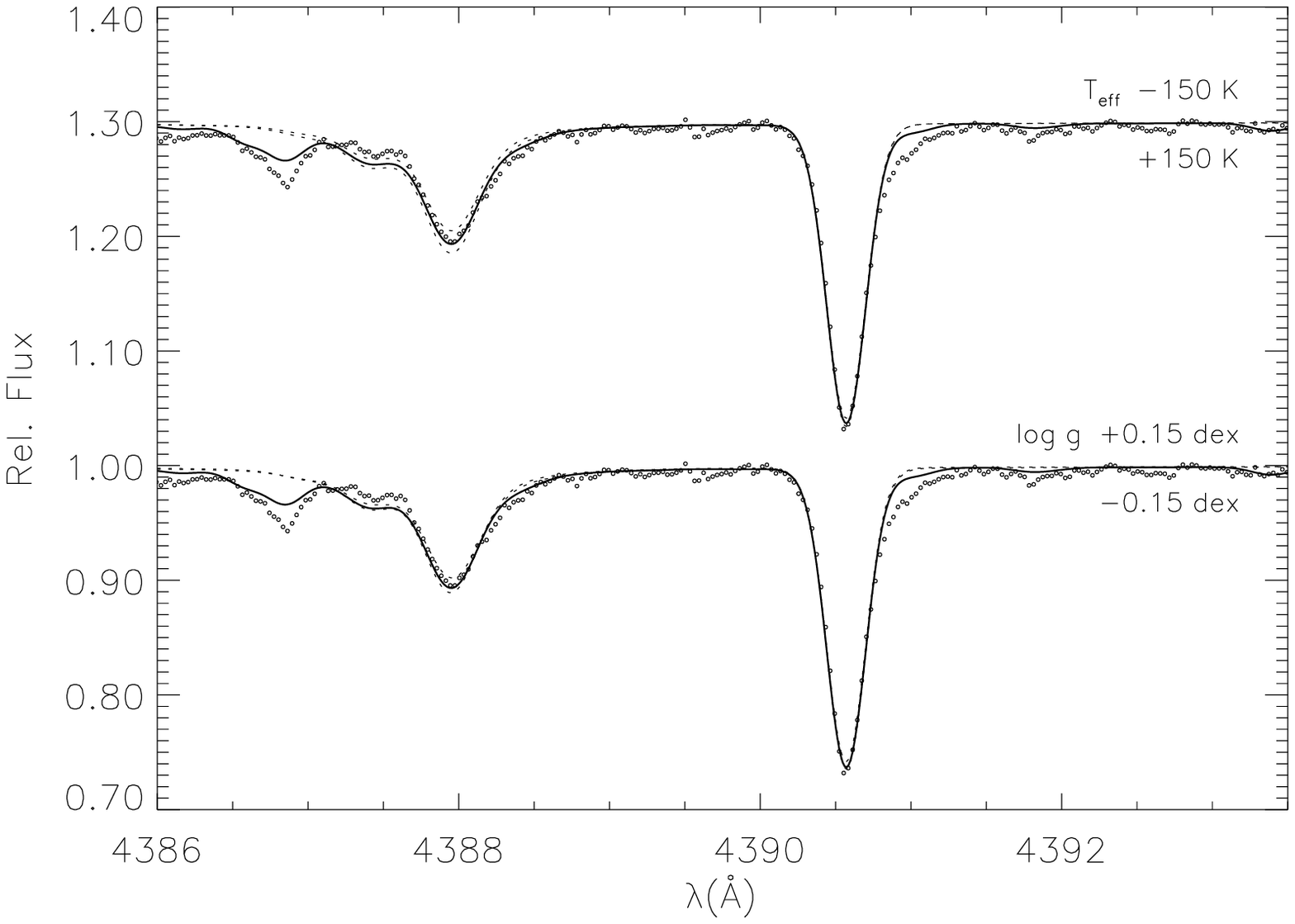}
\end{center}
\caption[]{
Temperature and gravity dependence of Mg\,{\sc i} (top) and Mg\,{\sc ii} 
(bottom) of $\eta$\,Leo. Results from the NLTE computations for the final
stellar parameters ({\it full line}) are compared with synthetic spectra for
modified parameters ({\it dotted lines}), as indicated, against observation
({\it dots}); from \cite{przy02}
}
\label{ftg2}
\end{figure}

\begin{figure}[ht]
\begin{center}
\includegraphics[width=.7\textwidth]{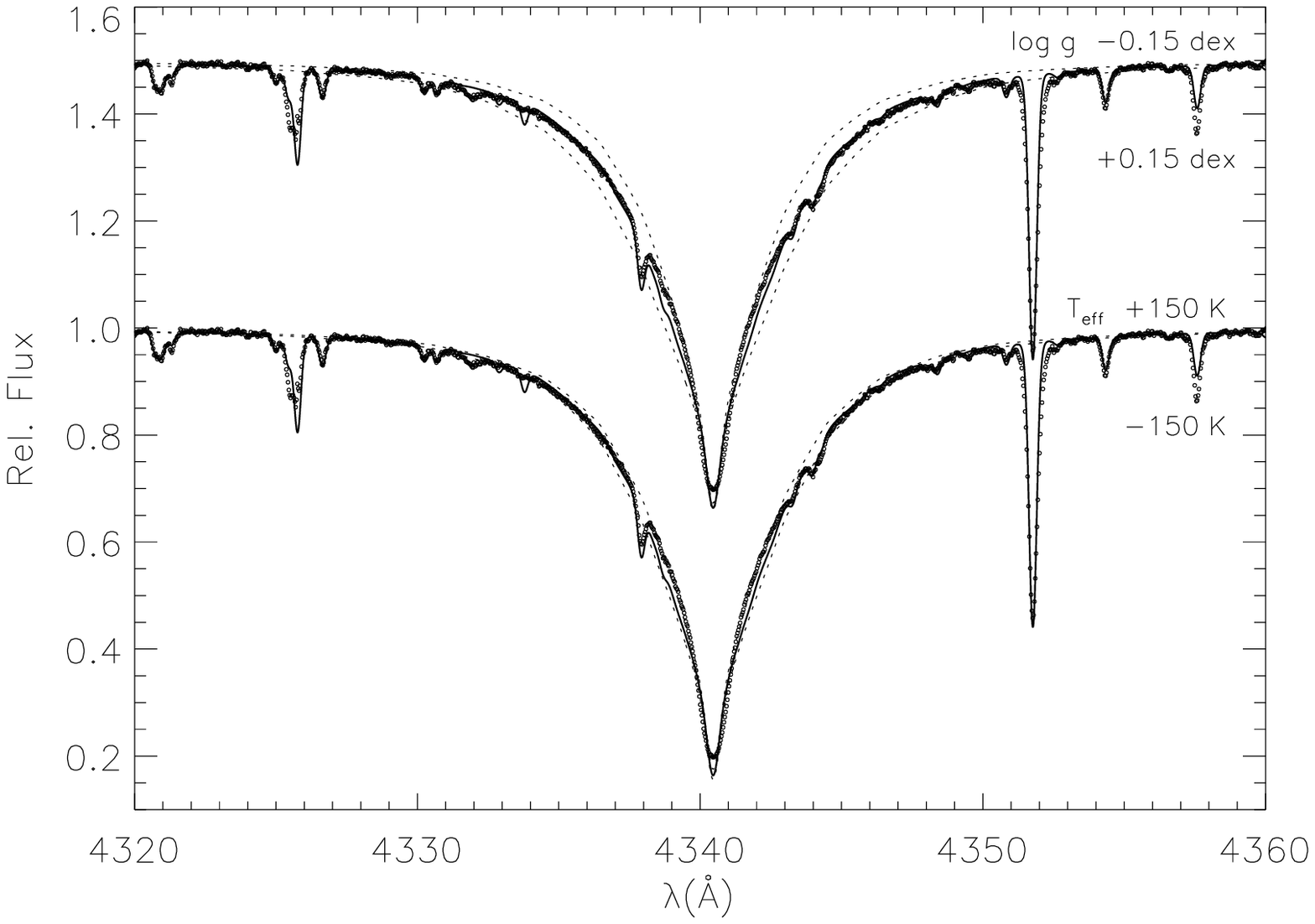}
\end{center}
\caption[]{
Temperature and gravity dependence of H$_{\gamma}$ of $\eta$\,Leo. See
Fig.~\ref{ftg2} for further annotations; from \cite{przy02}
}
\label{ftg3}
\end{figure}

For A-supergiants the technique has been pioneered by \cite{venn95}. Most
recent ex\-amples for applications are \cite{przy02}, \cite{venn99}, 
\cite{venn01a}, \cite{venn01b}. Examples are given in 
Figs.~\ref{ftg1}--\ref{ftg3}. 

The accuracy in the determination of T$_\mathrm{eff}$ and $\log g$, which can 
be achieved when using spectra of high S/N and sufficient resolution is
astonishing.\linebreak
$\Delta$T$_\mathrm{eff}$/T$_\mathrm{eff}$\,$\sim$\,0.01 and 
$\Delta\log g$\,$\sim$\,0.05 are realistic values.

\subsection{Chemical Composition}

\begin{figure}[ht]
\begin{center}
\includegraphics[width=.55\textwidth,angle=90]{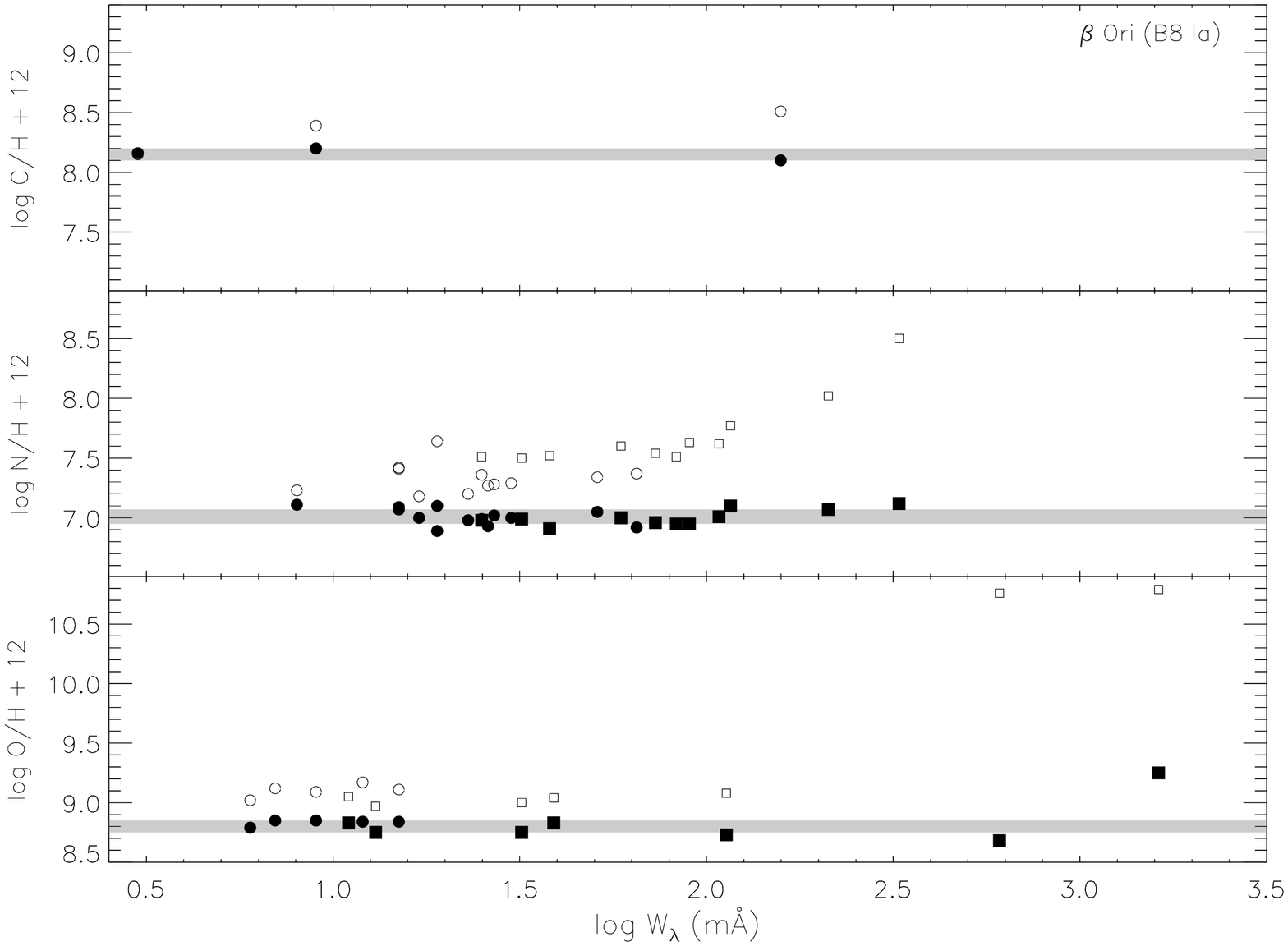}\\%[-.1cm]
\includegraphics[width=.55\textwidth,angle=90]{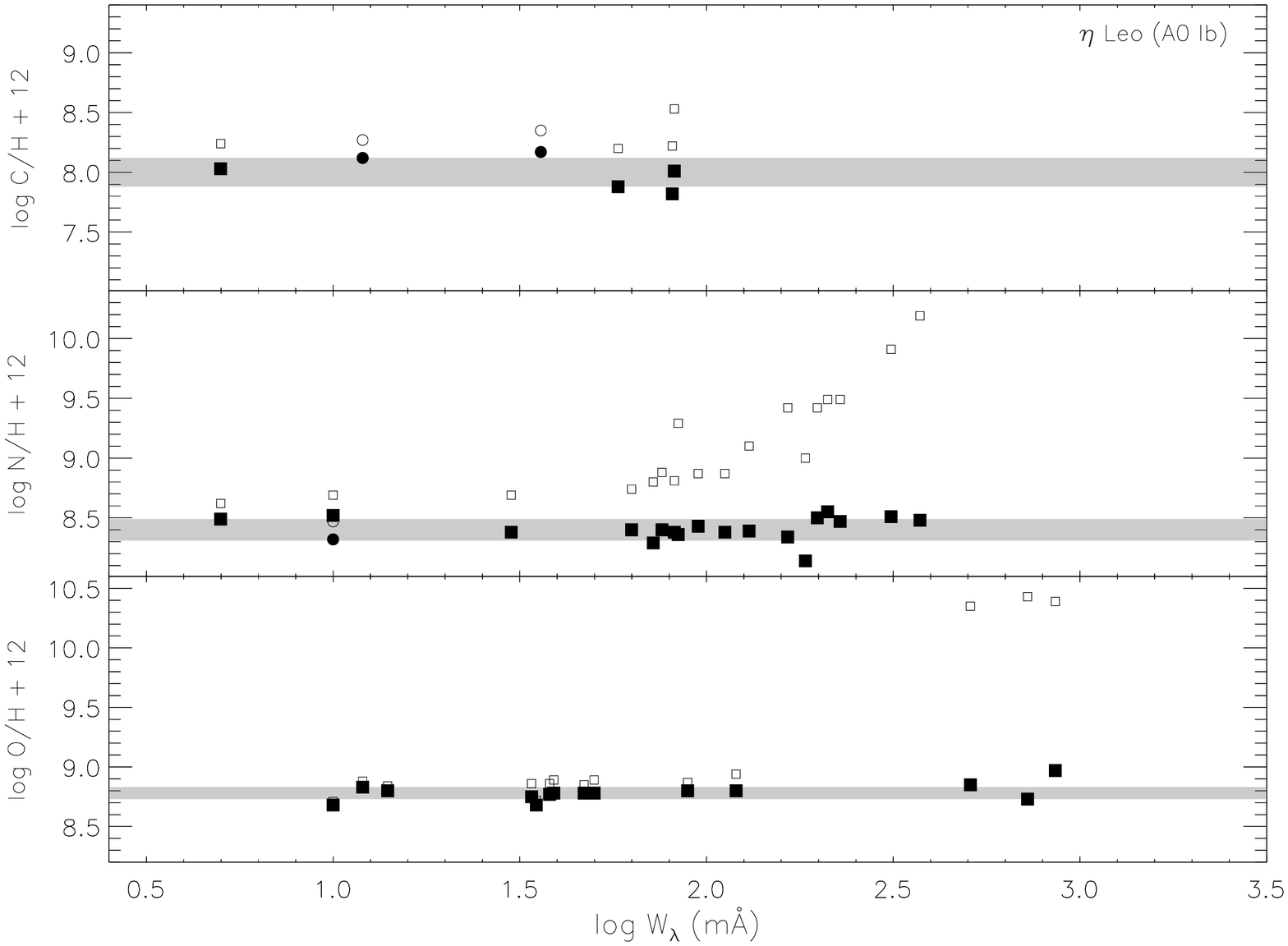}
\end{center}
\caption[]{Element abundances derived from individual spectral lines of CNO
plotted as a function of equivalent width. Top: $\beta$\,Ori, bottom: 
$\eta$\,Leo. Open symbols refer to LTE calculations for the line formation, whereas
solid symbols show the results of NLTE radiative transfer, for neutral ({\it
boxes}) and ionized species ({\it circles}). It is evident
that LTE fails badly, in particular, for stronger lines. The NLTE results
are remarkably consistent;
from \cite{przy03}
}
\label{abu1}
\end{figure}

The development of very detailed model atoms and using new and
very accurate atomic data, \cite{ip},~\cite{op}, has led to an
enormous improvement of the precision to which elemental abundances even in
extreme blue supergiants can be determined \cite{przy00}, \cite{przy01a},
\cite{przy01b}, \cite{przy01c}, \cite{przy02}, \cite{venn01a},
\cite{venn01b}. On the average, the uncertainties are now reduced to 
0.1\,dex in the abundance relative to hydrogen.
Figure~\ref{abu1} displays a nice example for the fit of the equivalent widths
of CNO lines in blue supergiants. 

\begin{figure}[ht]
\begin{center}
\includegraphics[width=.67\textwidth]{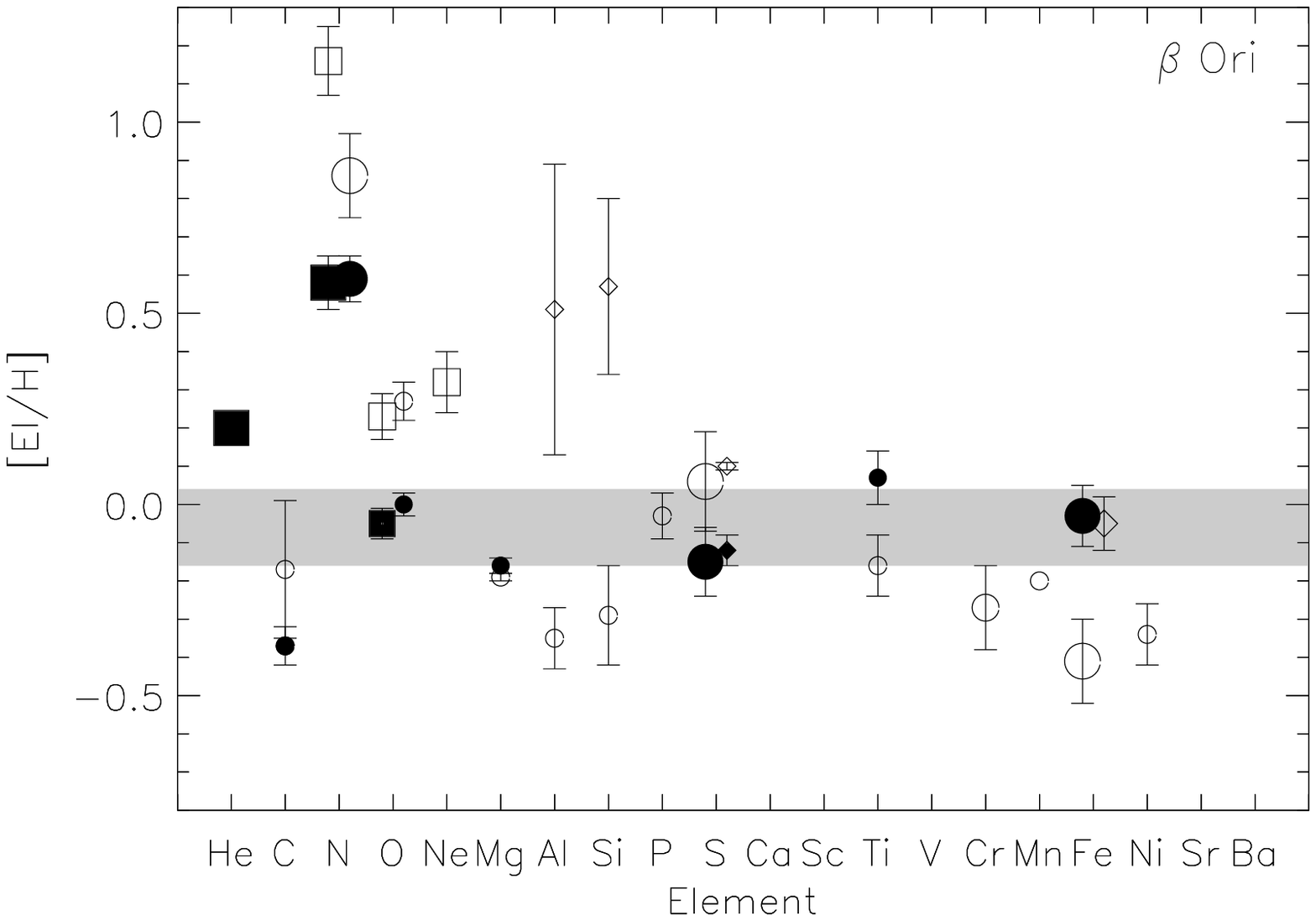}\\[-5mm]
\includegraphics[width=.67\textwidth]{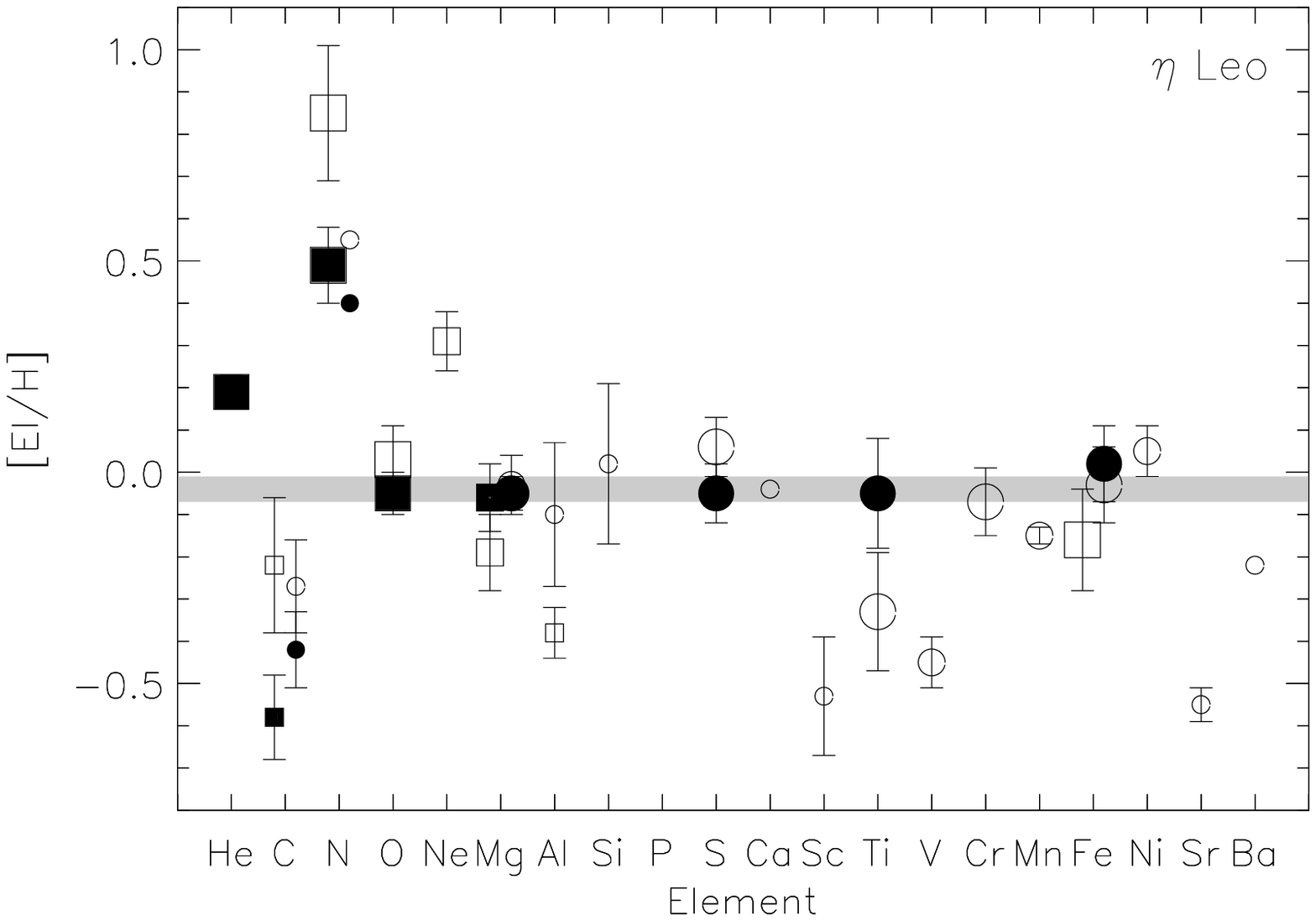}
\end{center}
\caption[]{
Abundance pattern determined for the two Milky Way supergiants $\beta$\,Ori and
$\eta$\,Leo, relative to the solar standard \cite{grev98} on a logarithmic 
scale. NLTE ({\it filled symbols}) and LTE abundances ({\it open
symbols}) for neutral ({\it boxes}), single-ionized ({\it circles}) and
double-ionized ({\it diamonds}) species. The symbol size codes the number of
spectral lines analyzed. Error bars represent 1$\sigma$-uncertainties from
the line-to-line scatter. The grey shaded area marks the deduced stellar
metallicity within 1$\sigma$-errors. The NLTE computations reveal a striking
similarity to the solar abundance distribution, except for the light
elements which have been affected by mixing with nuclear-processed matter; 
from \cite{przy02}
}
\label{abu2}
\end{figure}

\begin{figure}[ht]
\begin{center}
\includegraphics[width=.67\textwidth]{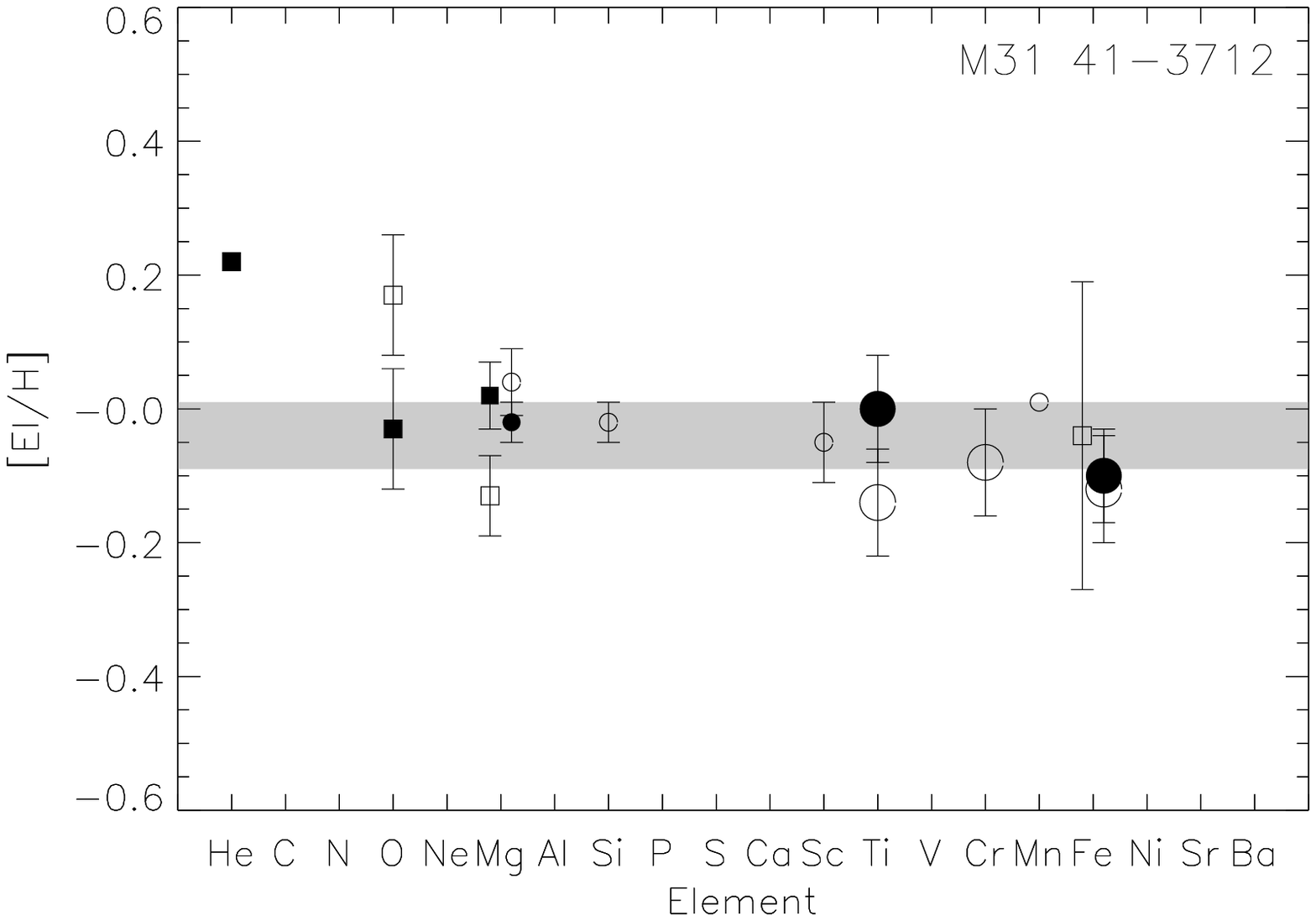}\\[-5mm]
\includegraphics[width=.67\textwidth]{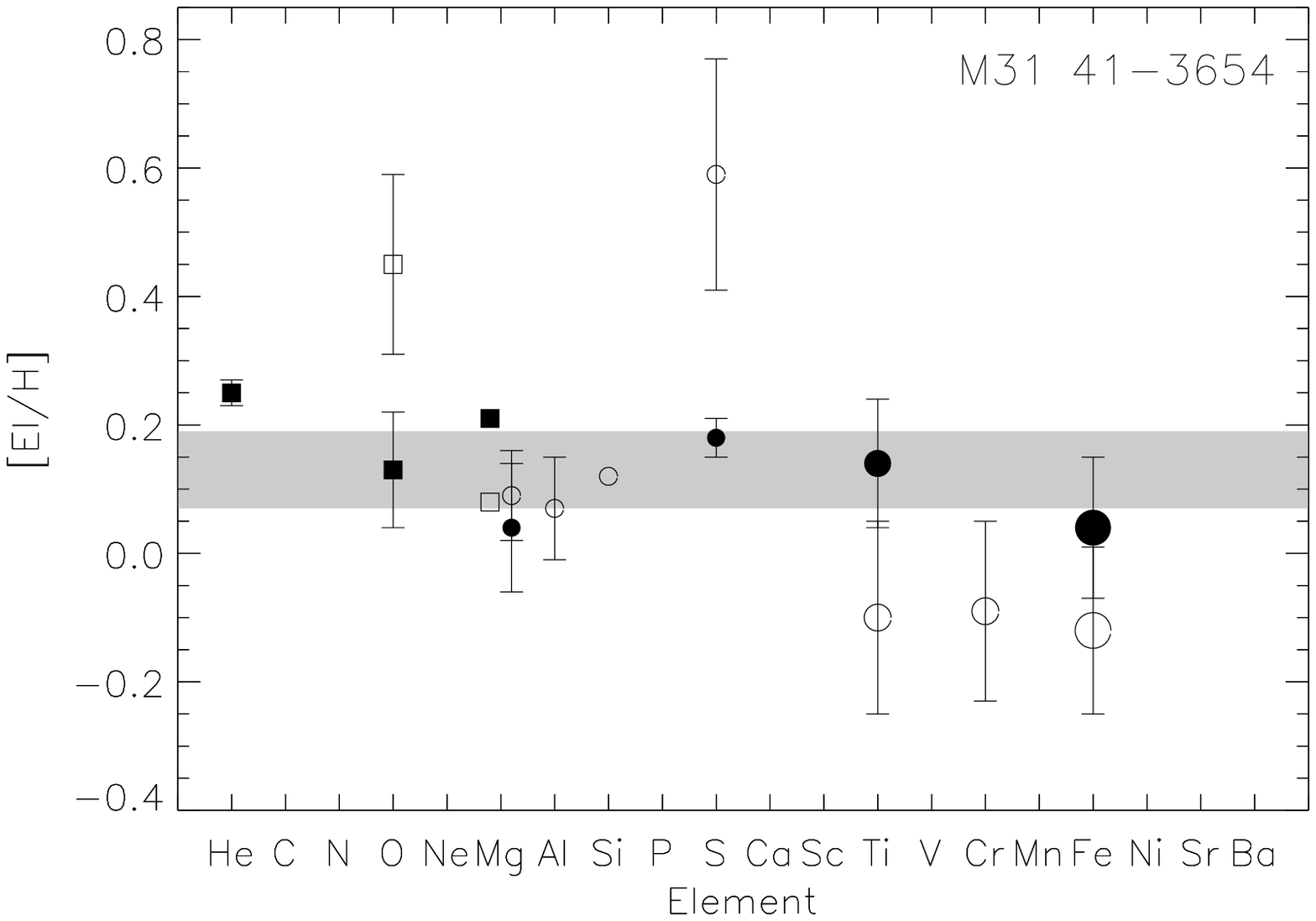}
\end{center}
\caption[]{
Abundance pattern of the two M\,31 A-supergiants 41-3712 and 41-3654. See
Fig.~\ref{abu2} for further annotations; from \cite{przy02}
}
\label{abu3}
\end{figure}

The amount of information about chemical elements is impressive.
Figure~\ref{abu2} gives an overview about the chemical elements the
abundances of which can be determined from the optical spectra of blue
supergiants. Figure~\ref{abu2} shows characteristic abundance patterns, as
they can be derived for supergiants in the Milky Way and Fig.~\ref{abu3}
displays results for two M\,31 supergiants.

\subsection{Stellar Wind Properties}

In principle, two types of lines are formed in a stellar wind,
P-Cygni profiles with a blue absorption trough and a red emission peak 
and pure emission profiles. The difference is caused by the re-emission 
process after
the photon has been absorbed  within the line transition. If the photon
is immediately re-emitted by spontaneous emission, then we have the case of
line scattering with a source function proportional to the geometrical
dilution of the radiation field and a P-Cygni profile will result. If the
re-emission occurs as a result of a different atomic process, for instance
after a recombination of an electron into the upper level or after a 
spontaneous decay of a higher level into the upper level or after a collision,
 then the line source function will possibly not dilute and
may roughly stay constant as a function of radius so that an emission line 
results. Typical examples for P-Cygni profiles are UV resonance transitions
connected with the ground level, whereas excited lines of an ionization stage
into which frequent recombination from higher ionization stages occurs will
produce emission lines. H$_{\alpha}$ in O-stars and early B-supergiants is a 
typical example for the latter case.
However, for late B- and
A-supergiants, when Ly$_{\alpha}$ becomes severely optically thick and the 
corresponding transition is in detailed balance, the first excited level of 
hydrogen becomes the effective groundstate and H$_{\alpha}$ starts to behave
like a resonance line showing also the shape of a P-Cygni profile (for a more
comprehensive discussion of the line formation process in winds see 
\cite{kud88}, \cite{kud98} and the most recent review \cite{kud2000}). 

Terminal velocities can be determined very precisely from the blue edges of
P-Cygni profiles and the red emission wings, normally with an accuracy of 5
to 10 percent (but see \cite{kud2000} for details). In addition,
H$_{\alpha}$ profiles normally allow for a very accurate (20 percent) 
determination of 
mass-loss rates in all cases of O, B, and A-supergiants \cite{puls96},
\cite{kud99}, \cite{mccarthy97}, but see \cite{kud2000} and \cite{puls2002}
for details and problems.

\begin{figure}[ht]
\begin{center}
\includegraphics[width=.6\textwidth]{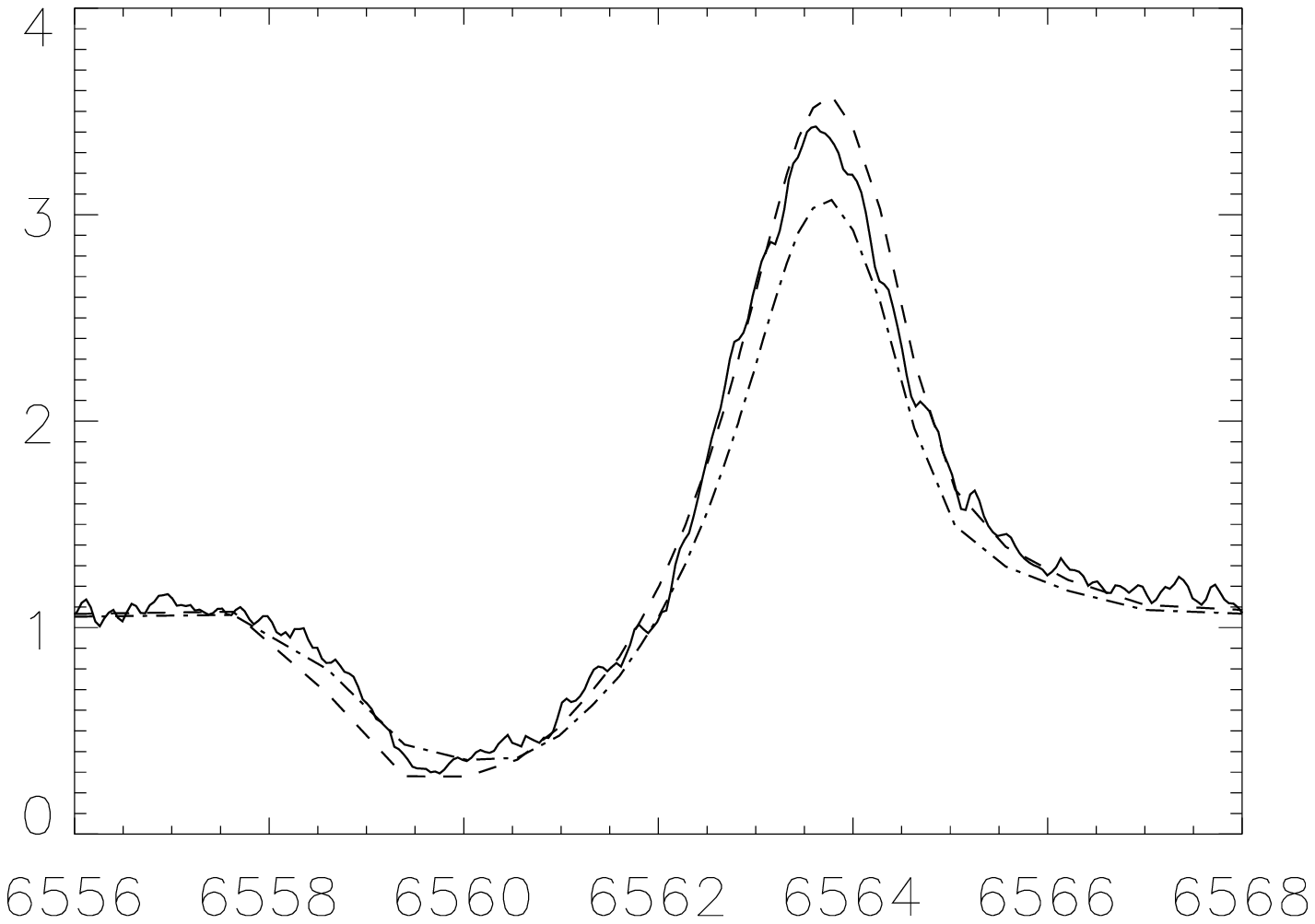}\\[-1cm]
%\hfill
\includegraphics[width=.6\textwidth]{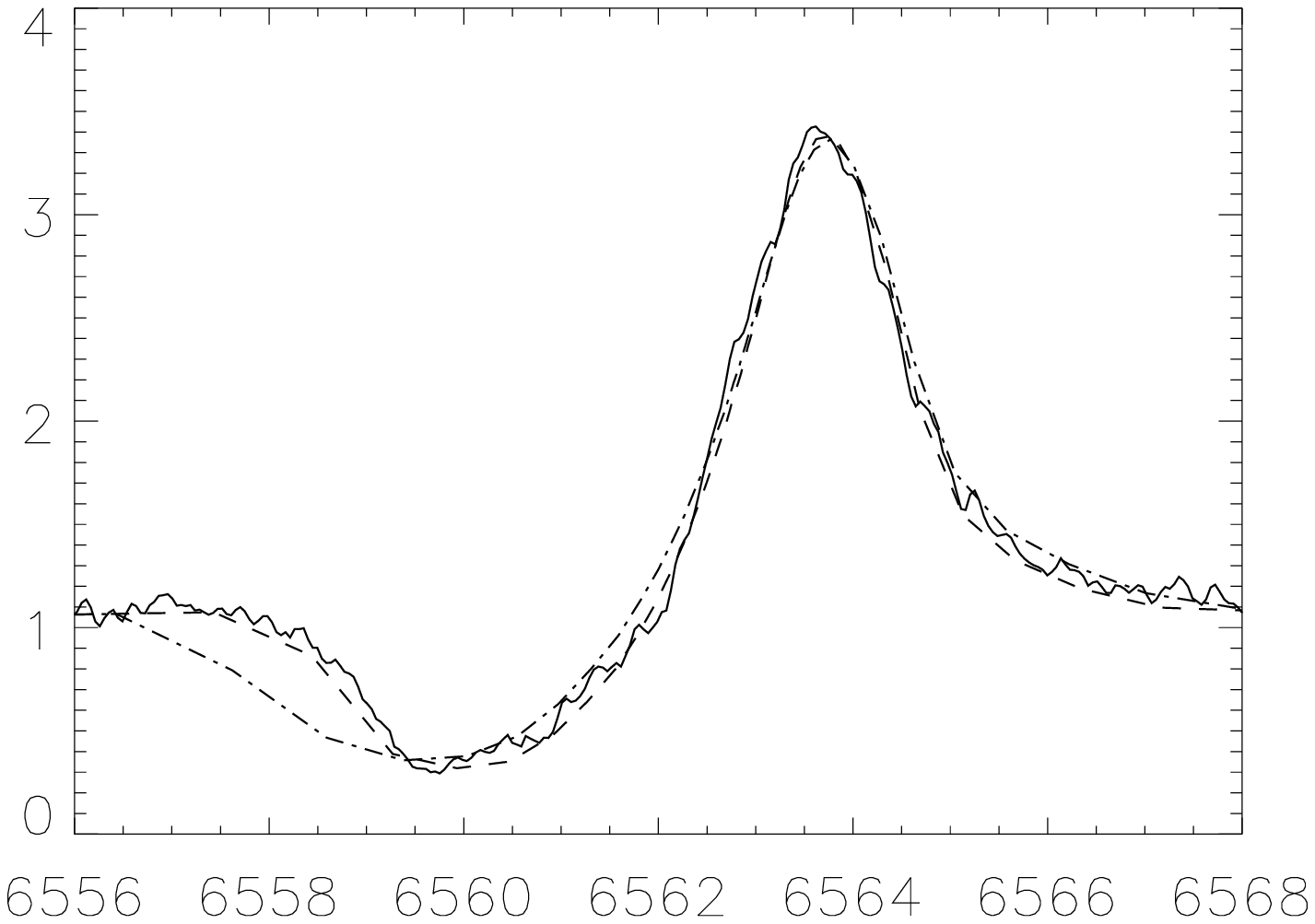}
\end{center}
\caption[]{Top: The influence of the mass-loss rate $\dot M$ on the
           H$_{\alpha}$ profile of the M\,31 A-supergiant 41-3654. 
	   Two models with $\dot M = 1.65$ and
           $2.15 \times 10^{-6}$\,M$_\odot$\,yr$^{-1}$ ({\it dashed-dotted}, 
	   {\it dashed}) 
           and otherwise identical parameters are shown superimposed on
           the observed profile. Bottom: The determination of $v_\infty$. 
	   Two models
           with $v_\infty = 200$ and $250\,$km\thinspace s$^{-1}$
           ({\it dashed}, {\it dashed-dotted}) and $\dot M$ adopted to fit the 
           height of the emission peak are shown superimposed to the observed 
           profile. All other parameters are identical; 
	   from~\cite{mccarthy97}
}
\label{awind1}
\end{figure}

Figure~\ref{awind1} gives an impression about the accuracy of the stellar wind
spectral diagnostics for A-supergiants.

\subsection{Spectral Resolution}

For extragalactic applications beyond the Local Group spectral resolution
becomes an issue. The important points are the following. Unlike the case
of late type stars, crowding and blending of lines is not a severe problem
for hot massive stars, as long as we restrict our investigation to the visual
part of the spectrum. In addition, it is important to realize that massive
stars have angular momentum, which leads to usually high rotational velocities.
Even for A-supergiants, which have already expanded their radius
considerably during their evolution and, thus, have slowed down their
rotation, the observed projected rotational velocities are still on the order
of 30\,km\,s$^{-1}$ or higher. This means that the intrinsic full half-widths 
of metal lines are on the order of 1\,{\AA}. In consequence, for the detailed
studies of supergiants in the Local Group a resolution of 25,000 sampling a 
line with five data points is ideal. This is indeed the resolution, which has
been applied in most of the work referred to in the previous sections.

However, as we have found out empirically \cite{przy02}, degrading the
resolution to 5,000 (FWHM = 1\,{\AA}) has only a small effect on the accuracy
of the diagnostics, as long as the S/N remains high (i.e. 50 or better).
Even for a resolution of 2,500 (FWHM = 2\,{\AA}) it is still possible to
determine T$_\mathrm{eff}$ to an accuracy of 2 percent, $\log g$ to 0.05 dex and
individual element abundances to 0.1 or 0.2\,dex. 

References
\cite{bresolin01}, \cite{bresolin02}, \cite{bresolinWN}, \cite{bres03} and
\cite{kud2003} have used FORS at the VLT with a resolution of 1,000 
(FWHM = 5\,{\AA}) to study blue supergiants far beyond the Local Group. The
accuracy in the determination of stellar properties at this rather low
resolution is still remarkable. The effective temperature is accurate to
roughly 4 percent and the determination of gravity remains unaffected and is
still good to 0.05\,dex (an explanation will be given in Sect.~4). However,
at this resolution it becomes difficult to determine abundance patterns of
individual elements (except for emission line stars, see \cite{bresolinWN})
and, thus, one is restricted to the determination of the overall metallicity 
which is still accurate to 0.2\,dex.

We can conclude that
an optimum for extragalactic work is a resolution of 2,500 (FWHM = 2\,{\AA}).
Blue supergiants are bright enough even far beyond the Local Group to allow the
achievement of high S/N with reasonable exposure times at 8m-class telescopes
with efficient multi-object spectrographs at this resolution, which then
makes it possible to determine stellar parameters and chemical composition
with sufficient precision. This resolution is also good enough to determine
stellar wind parameters from the observed H$_{\alpha}$ profiles, since the
stellar wind velocities (and therefore the corresponding line widths) are
larger than 150\,km\,s$^{-1}$.

\section{The Wind Momentum--Luminosity Relationship}

The concept of the Wind Momentum--Luminosity Relationship (WLR) has been
introduced by \cite{kud95} and \cite{puls96}. It starts
from a very simple idea. The winds of blue supergiants
are initiated and maintained by the absorption of photospheric radiation and
the photon momentum connected to it. Thus, the mechanical momentum flow of
a stellar wind $\dot{M} v_{\infty}$ should be a function of the photon
momentum rate $L/c$ provided by the stellar photosphere and interior

\begin{equation}
\dot{M} v_{\infty} = f(L/c)\;.
\end{equation}

If this is true and if we are able to find this function $f$, this would
enable us to determine directly stellar luminosities from the stellar wind
by using the inverse relation
\begin{equation}
L = f^{-1}(\dot{M} v_{\infty})\;.
\end{equation}
 
In other words, by measuring the rate of mass-loss and the terminal velocity
directly from the spectrum we would be able to determine the luminosity of
a blue supergiant. This is an exciting perspective, because it would give us
a completely new, purely spectroscopic tool to determine stellar distances.
Quantitative spectroscopy would yield $T_\mathrm{eff}$, gravity, abundances,
intrinsic colours, reddening, extinction, $\dot{M}$ and $v_{\infty}$.
With the luminosity from the above relation one could then compare with
the dereddened apparent magnitude to derive a distance.

In the previous section, we already demonstrated how $\dot{M}$ and 
$v_{\infty}$ can be determined from the spectrum with high precision.
Thus, deriving the theoretical relationship and confirming and calibrating
it observationally will enable us to introduce a new distance determination
method. An accurate analytical solution of the hydrodynamic equations of
line driven winds has been provided by \cite{kud89}. These solutions were
then used by \cite{kud95} and \cite{puls96} to exactly derive the relationship.
Here, we apply a simplified approach, see also~\cite{kud98}, which is not
exact but gives insight in the underlying physics.

\begin{figure}[t]
\begin{center}
\includegraphics[width=.365\textwidth,]{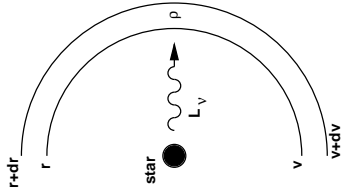}
\end{center}
\caption[]{
Sketch of a blue supergiant irradiating its own stellar wind envelope.
$L_{\nu}$ is the spectral luminosity at frequency $\nu$. $v$ is the wind
outflow velocity at radius $r$ and $\varrho$ is the mass density
}
\label{wlrsy}
\end{figure}

We assume that the wind is stationary and spherical symmetric and obeys
the equation of continuity
 \begin{equation}
\dot{M} = 4\pi r^{2} \varrho(r) v(r)\;.
\end{equation}

Then we consider the star as a point source of
photons irradiating and accelerating its own stellar wind
(see Fig.\,\ref{wlrsy}) and calculate the amount of photon momentum
absorbed by one spectral line
\begin{equation}
\fre{L}{\fri{c}} \fre{L_{\nu_{i}} (1-e^{-\tau_{i}}) \D \nu^\mathrm{width}}
{\fri{L}} = \fre{L}{\fri{c^{2}}} \fre{\nu_{i} L_{\nu_{i}}} {\fri{L}}
(1-e^{-\tau_{i}}) \D v\;.
\end{equation}
 The first factor on the left side gives the total photon momentum rate
provided by the star, the second describes the fraction absorbed by one
spectral line in an outer shell of thickness $\D r$.
$\tau_{i}$ is the optical thickness of such an outer shell in the
line transition $i$. $L_{\nu_{i}} \D \nu^\mathrm{width}$ is the stellar spectral
luminosity at the frequency of line $i$ multiplied by the spectral width of
the line.
This luminosity can in principle be absorbed by the line if it is entirely
optically thick (i.e. $\tau_{i} \gg 1$). However, depending on the optical
thickness only the fraction $(1-e^{-\tau_{i}})$ is really absorbed. If we
are in the supersonic part of the wind, then the spectral width
$\D \nu^\mathrm{width}$ is not determined by the thermal motion of the ions
but rather by the increment of the velocity outflow $\D v$ via the Doppler
formula
\begin{equation}
\D \nu^\mathrm{width} = \nu_{i} \fre{\D v}{\fri{c}}\;,
\end{equation}
which leads to the right hand side of (4).

After calculation of the photon momentum absorbed by a single spectral line we
can consider the momentum balance in the stellar wind. The photon momentum
absorbed by all lines will just be a sum over all lines $i$ of the
expression shown on the right hand side of (4). Almost all of this
absorbed momentum will be transformed into gain of mechanical stellar wind
flow momentum $\dot{M}\D v$ of the outer shell except the fraction
$g(r)\D M_{r}$, which is the momentum required to act against the gravitational
force ($g(r)=GM_{\ast}/r^{2}$ is the local gravitational acceleration and
$\D M_{r}=\varrho 4\pi r^{2}\D r$ is the mass within the spherical stellar wind
shell)
\begin{equation}
\dot{M}\D v = \fre{L}{\fri{c^{2}}} \sum_{i} \fre{\nu_{i}L_{\nu_{i}}} {\fri{L}}
(1-e^{-\tau_{i}})dv - G\fre{M{\ast}(1-\Gamma)}{\fri{r^{2}}}\varrho 4\pi
r^{2}\D r\;.
\end{equation}
Note that this momentum balance also includes the photon momentum
transfer by Thomson scattering, which leads to the correction factor
$(1-\Gamma)$ in the local gravitational acceleration.
 
Now, the important next step is to deal with the sum over all lines in the
above
momentum balance. Here, the complication arises from the term in parentheses
containing the local optical depth $\tau_{i}$ of each line. $\tau_{i}$ will not
only be different for each of the thousands of lines driving the wind, it will
also vary through the stellar wind as a function of radius. On the other hand,
one of the enormous simplifications in supersonically expanding envelopes
around stars is that the optical thickness is well described by (see
for instance \cite{kud88})
\begin{equation}
\tau_{i} = k_{i} \kappa_\mathrm{Thom} \fre{v_\mathrm{therm}}{\fri{\D v/\D r}}\;,
\end{equation}
where $v_\mathrm{therm}$ is the thermal velocity of the ion and $k_{i}$
is the (dimensionless) \emph{line strength} defined as
\begin{equation}
k_{i} = \fre{\kappa_{i}}{\fri{\kappa_\mathrm{Thom}}}\;,
\end{equation}
i.e. the opacity of line $i$
\begin{equation}
\kappa_{i} = \fre{1}{\fri{\Delta\nu_\mathrm{Dopp}}} \fre{\pi e^{2}}
{\fri{m_\mathrm{e}c}}n_{l}f_{lu}\left(1-\fre{n_{u}}{\fri{n_{l}}}\fre{g_{l}}{\fri{g_{u}}}\right)
\end{equation}
in units of the continuous Thomson scattering opacity or in units of the
local density
\begin{equation}
\kappa_\mathrm{Thom} = n_\mathrm{e}\sigma_\mathrm{e} = \varrho s_\mathrm{e}\;.
\end{equation}
$\sigma_\mathrm{e}$ is the cross section for Thomson scattering of photons on free
electrons, $n_\mathrm{e}$ the local number density of free electrons. In a hot plasma
with hydrogen as the main constituent mostly ionized and with
$Y_\mathrm{He}=n_\mathrm{He}/n_\mathrm{H}$ and $I_\mathrm{He}$ the number of electrons
provided per helium nucleus we have ($m_\mathrm{H}$ is the mass of the hydrogen
atom)
\begin{equation}
s_\mathrm{e} = \fre{1+I_\mathrm{He}Y_\mathrm{He}}{\fri{1+4Y_\mathrm{He}}}
\fre{\sigma_\mathrm{e}}{\fri{m_\mathrm{H}}}\;.
\end{equation}
 
Thus, if the degree of ionization of helium is roughly constant as a function
of radius $r$ in the wind, $s_\mathrm{e}$ is also constant and we have
\begin{equation}
\tau_{i} = k_{i} s_\mathrm{e} \varrho (r) \fre{v_\mathrm{therm}}{\fri{\D v/\D r}}\;,
\end{equation}
with the line strength $k_{i}$ proportional to oscillator strength
$f_{lu}$, wavelength $\lambda_{i}$ and the occupation number $n_{l}$ of the
lower level divided by the mass density $\varrho$
\begin{equation}
k_{i} \propto \fre{n_{l}}{\fri{\varrho}} f_{lu} \lambda_{i}\;.
\end{equation}

Thus, the line strength is roughly independent of the depth in the atmosphere
and is determined by atomic physics ( $f_{lu}$, $\lambda_{i}$) and atmospheric
thermodynamics ($n_{l}/\varrho$). Since $\varrho$ and $\D v/\D r$ vary strongly through
the wind a line can be optically thick in deeper layers
\begin{equation}
\tau_{i} \gg 1 \Longrightarrow (1-e^{-\tau_{i}})\D v \approx \D v
\end{equation}
and can become optically thin further out
\begin{equation}
\tau_{i} \ll 1 \Longrightarrow (1-e^{-\tau_{i}})\D v \approx k_{i}\varrho \D r\;.
\end{equation}

For the calculation of the photon momentum transfer this means that
line contributions can have an entirely different functional form depending
on the optical thickness of the lines.

This problem can be solved in a very elegant way by introducing a \emph{line
strength distribution function}
\begin{eqnarray*}
n(k,\nu)\D \nu \D k & = & \mbox{number of lines with $\nu_{i}$ from 
($\nu$,\,$\nu$\,$+$\,$\D \nu$)}\\
& & \mbox{and with $k$ from ($k$,\,$k$\,$+$\,$\D k$)}\;.
\end{eqnarray*}

Since the modern hydrodynamic model atmosphere codes (see Sect.~2) contain
atomic data and occupation numbers for millions of lines in NLTE, we
can investigate the physics of the line strength distribution function. As it
turns out, \cite{spring97}, \cite {kud2000}, \cite{puls2000} and 
\cite{kud02},
the distribution in line strengths -- to a very good approximation -- obeys a 
power law
\begin{equation}
n(k,\nu)\D \nu \D k = g(\nu)\D \nu k^{{\bf \alpha}-2} \D k\;, 1 \leq k \leq \infty
\end{equation}
independent (to first order) of the frequency. The exponent {\bf $\alpha$}
depends weakly on T$_\mathrm{eff}$ and varies (in the temperature range of OB-stars)
between
\begin{equation}
\alpha = 0.6 \ldots 0.7\;.
\end{equation}

$\alpha$ is mostly determined by the atomic physics and basically
reflects
the distribution function of the oscillator strengths. One can,
for instance, show
that for the hydrogen atom the distribution of the Lyman-series oscillator
strengths is a power law with exponent $\alpha=2/3$, see~\cite{kud98}. 
It is important to
realize that $\alpha$ is not a free parameter  but, instead, is well determined
from the thousands of lines taken into account in the model atmosphere
calculations. Examples for the power law dependence of line strengths are given
in \cite{kud98} or \cite{puls2000}.

With the line strength distribution function we can replace
the sum in the momentum balance by a double integral, which
is then analytically solved (see also \cite{kud88}):
\begin{equation}
\sum_{i} \fre{\nu_{i}L_{\nu_{i}}} {\fri{L}}(1-e^{-\tau_{i}})
\longrightarrow \int\limits_{0}^{\infty} \int\limits_{0}^{\infty}(1-e^{\tau (k)})
\fre{\nu L_{\nu}} {\fri{L}} n(k,\nu) \D \nu \D k
\longrightarrow N_{o} 
\left\{\frac{\D v/\D r}{\varrho}\right\}^{\alpha-1}\;.
\end{equation}

This means that the momentum transfer from photons to the stellar wind plasma
depends non-linearly on the gradient of the velocity field. The degree of the
non-linearity is determined by the steepness of the line strength distribution
function {\bf $\alpha$}. $N_{o}$ is proportional to the number of lines in the
line strength interval $1\leq k \leq \infty$.

We can now re-formulate the momentum balance to obtain
\begin{equation}
\dot{M}\D v = \fre{L}{\fri{c^{2}}}N_{o} \left\{ \frac{\D v/\D r}{\varrho}
\right\}^{\alpha-1}\D v - G\fre{M_{\ast}(1-\Gamma)}{\fri{r^{2}}}4\pi r^{2}\varrho \D r
\end{equation}
yielding a non-linear differential equation for the stellar wind velocity
(note that we replaced the density through the mass conservation equation)
\begin{equation}
r^{2}v\fre{\D v}{\fri{\D r}}=\fre{L}{\fri{\dot{M}^{{\bf
\alpha}}}}\fre{N_{o}}{\fri{c^{2}}}(4\pi)^{\alpha-1}\left\{ r^{2}
v\fre{\D v}{\fri{\D r}}\right\}^{ \alpha} - GM_{\ast}(1-\Gamma)
\end{equation}
which looks much more complicated than it really is. The solution is easy,
see~\cite{kud89}. We obtain $\dot{M}$ as the uniquely determined
eigenvalue of the problem
\begin{equation}
\dot{M} \propto L^{1/\alpha}\lbrace M_\ast(1-\Gamma)\rbrace^{1-1/\alpha}
\end{equation}
and a terminal velocity proportional to the escape velocity $v_\mathrm{esc}$
\begin{equation}
v_{\infty} \propto v_\mathrm{esc} \propto \left\{ G M_\ast
(1-\Gamma)/R_\ast\right\}^{1/2}\;.
\end{equation}
Combining the two yields the stellar wind momentum
\begin{equation}
\dot{M}v_{\infty} \propto \fre{1}{\fri{R_{\ast}^{1/2}}}L^{1/\alpha}
\lbrace M_\ast(1-\Gamma)\rbrace^{3/2-1/\alpha}\;,
\end{equation}
which -- as expected -- depends strongly on the luminosity but also on the
photospheric radius and the expression in the parentheses, which contains the
stellar mass and distance from the Eddington limit. It is this expression which
can vary significantly for different blue supergiants and, therefore, causes
the large scatter in the observed correlations of mass-loss
rates with
luminosity and terminal velocity with escape velocity as discussed previously.
 However, for the product of mass-loss rate
and terminal
velocity, the stellar wind momentum rate, the exponent of the term in brackets
should be -- thanks to the laws of atomic physics --
very close to zero, since $\alpha \approx 2/3$. This means that to
first order the wind momentum rate should be determined by
 \begin{equation}
{\bf \dot{M}v_{\infty} \propto \fre{1}{\fri{R_{\ast}^{1/2}}}
{\bf L}^{\bf 1/\alpha}}\;.
\end{equation}

\begin{figure}[t]
\begin{center}
\includegraphics[width=.39\textwidth,angle=90]{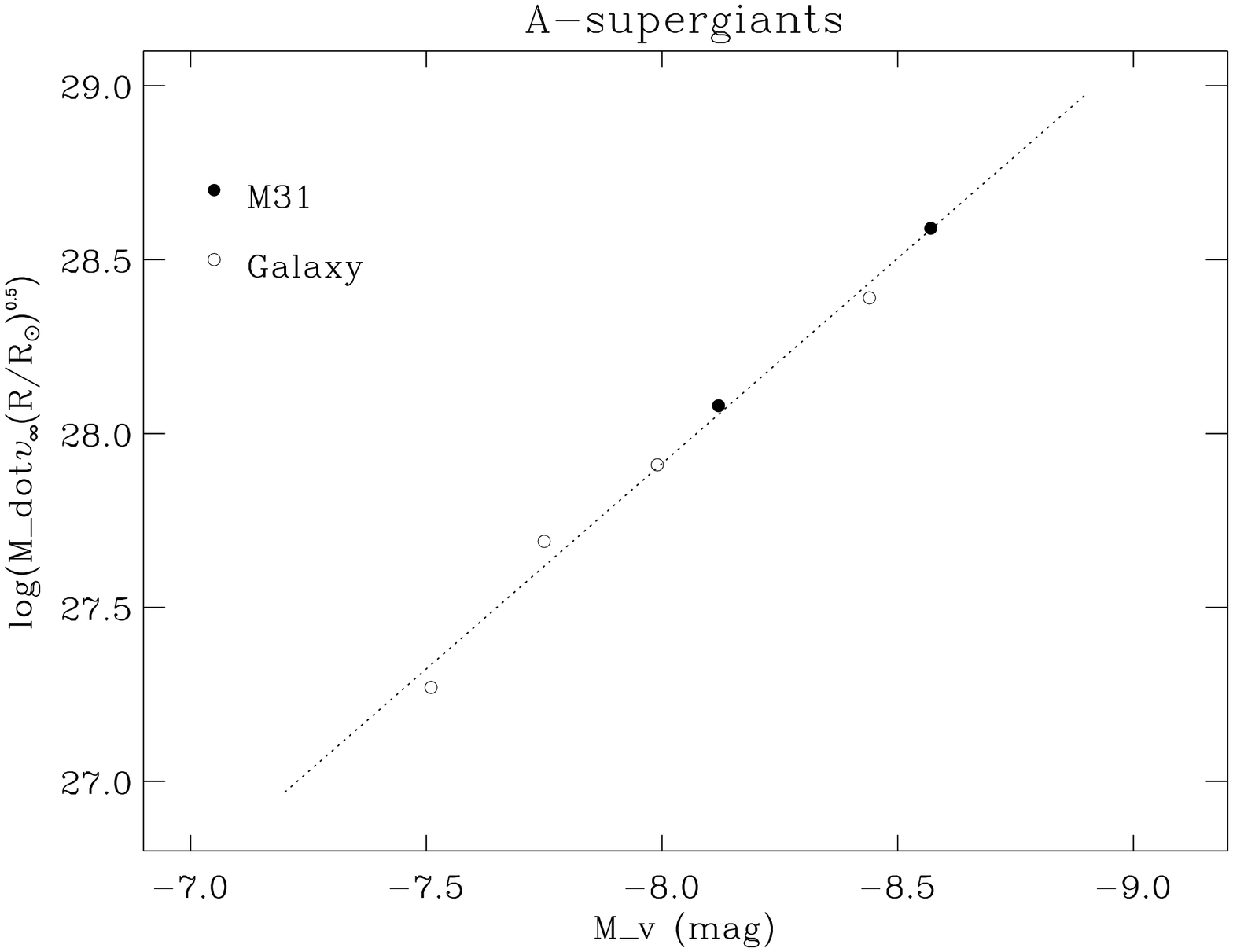}\\[-.1cm]
%\hfill
\includegraphics[width=.39\textwidth,angle=90]{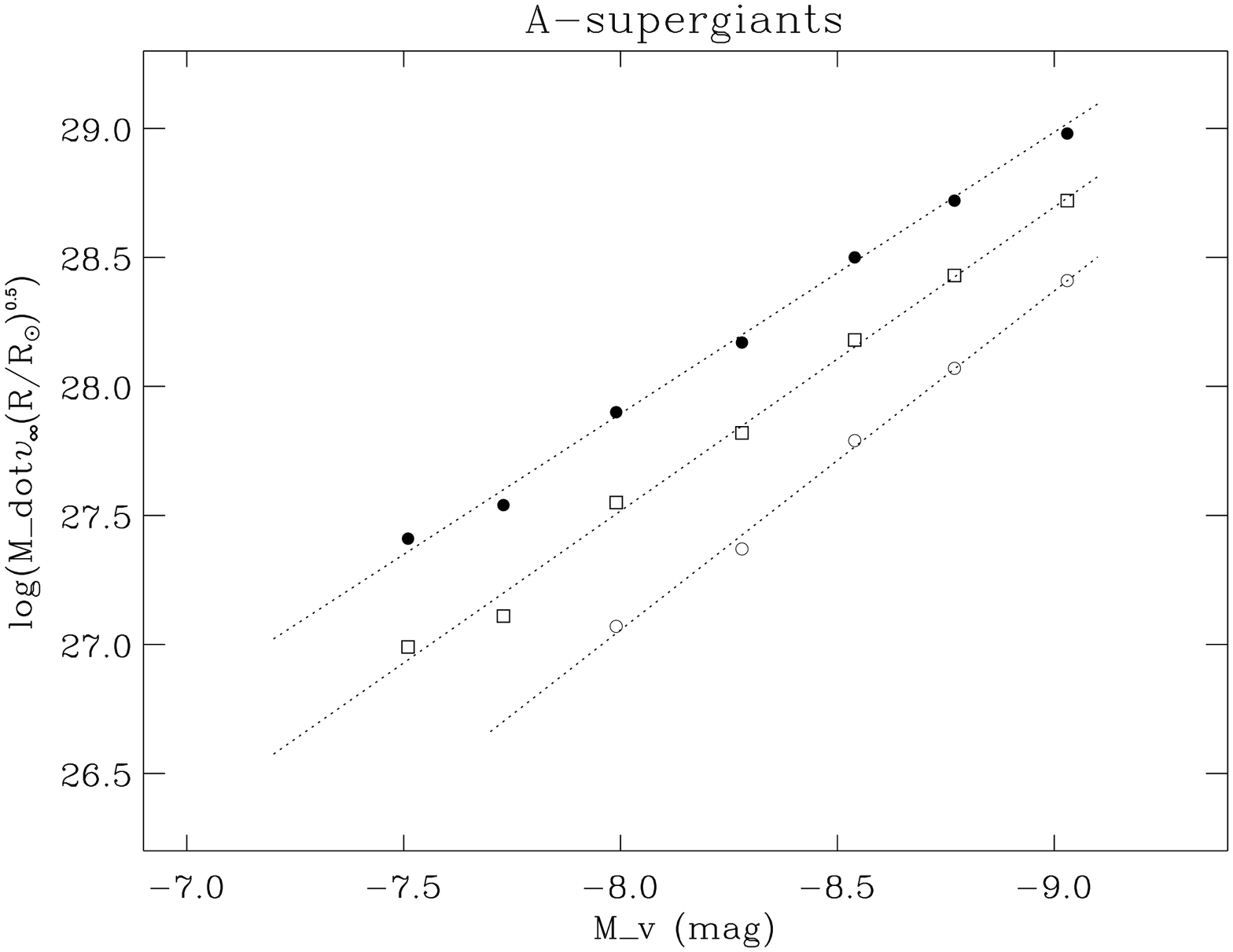}
\end{center}
\caption[]{The Wind Momentum -- Luminosity Relationship of A-supergiants.
Top: observed stellar wind momentum as a function of absolute magnitude for
objects in the Milky Way and M\,31; from~\cite{bresolin01}. Bottom: 
Theoretical calculations using
the stellar wind code by~\cite{kud02}, for solar metallicity 
({\it solid circles}),
0.4 solar ({\it open squares}) and 0.1 solar ({\it open circles})
}
\label{wlra}
\end{figure}

This is the \emph{Wind Momentum - Luminosity Relationship}. It predicts a
strong dependence of wind momentum rate on the stellar luminosity with an
exponent determined by the statistics of the strengths of the tens of
thousands of
lines driving the wind. It also contains a weak dependence on stellar radius
which comes from the fact that the stellar wind has to work against the
gravitational potential when accelerated by photospheric photons.

References \cite{puls96}, \cite{kud99} and \cite{kud2000} were the first to 
prove that the theoretically predicted WLR is really observed. O-stars, B and 
A-supergiants all follow this relationship. As to be expected, the
relationship depends on spectral types, since lines of different ionization
stages contribute to the mechanism of line driving at different spectra types.
F.~Bresolin, in this volume~\cite{bres03}, gives a detailed overview about
the most recent observational work on the WLR. Here, we only want to show
the recent best calibration for A-supergiants using (only) four Milky Way
and two M31 objects in Fig.\,\ref{wlra}. Although
a calibration based on merely six objects is only moderately convincing, we
are again encouraged by the small scatter over the remarkable range in
luminosity. Future calibration work will be crucial to establish the method
as an accurate distance indicator.

Also shown in Fig.~\ref{wlra} are recent theoretical stellar wind
calculations for A-supergiants (R.P. Kudritzki, in prep.), which are
based on the new radiation driven wind algorithm developed by \cite{kud02}.
The calculations provide a clear prediction about the metallicity dependence
of A-supergiant wind momenta in agreement with previous work discussed in
\cite{kud2000} and the recent work by~\cite{vink00} and~\cite{vink2001}.
F. Bresolin, in this volume~\cite{bres03}, will discuss most recent
extragalactic stellar wind diagnostics on blue supergiants and compare them 
with the model predictions.

\section{The Flux-Weighted Gravity--Luminosity Relationship}

It is an old idea that the strengths of the hydrogen Balmer lines and the
absolute luminosities of massive stars must be related (see \cite{bres03} in
this volume for an overview). The concept behind this idea is very simple.
Because of the effects of Stark broadening the Balmer lines in hot stars are
very sensitive to the number densities of electrons and protons in those
atmospheric layers, where the wings of the Balmer lines are formed. The
number densities, on the other hand, depend on the photospheric gravity as
the result of the hydrostatic equilibrium in stellar photospheres. Stellar 
gravities, however, reflect the evolution of massive stars away from the
ZAMS (where all gravities are roughly the same) towards lower effective
temperatures. They become smaller, as further the star evolves, and, if we
compare objects with exactly the same effective temperature, more
massive supergiant stars with higher luminosities are expected to have lower 
gravities than their counterparts with lower mass.

\begin{figure}[!ht]
\begin{center}
\includegraphics[width=.4\textwidth,angle=90]{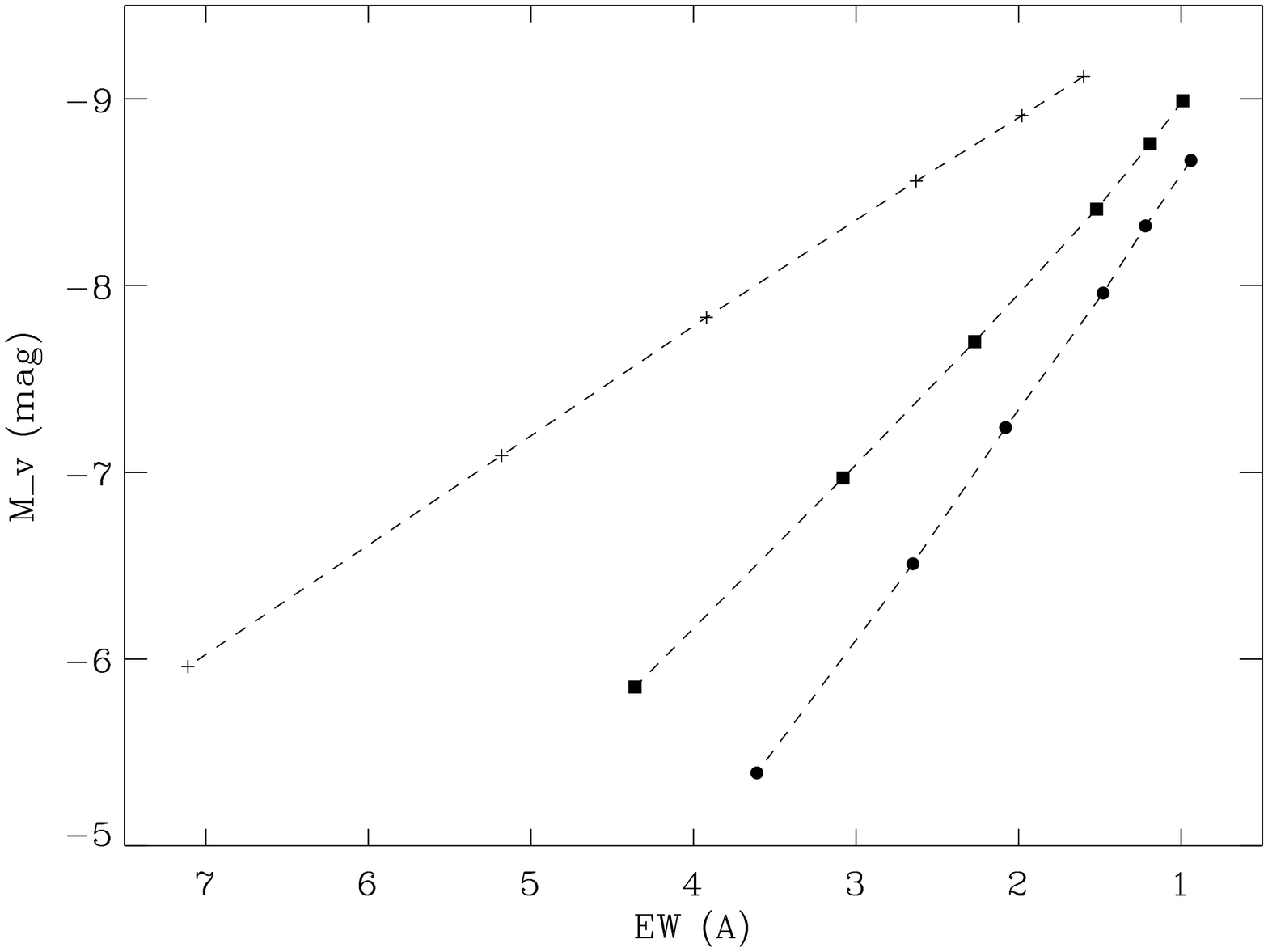}\\[-4mm]
\includegraphics[width=.4\textwidth,angle=90]{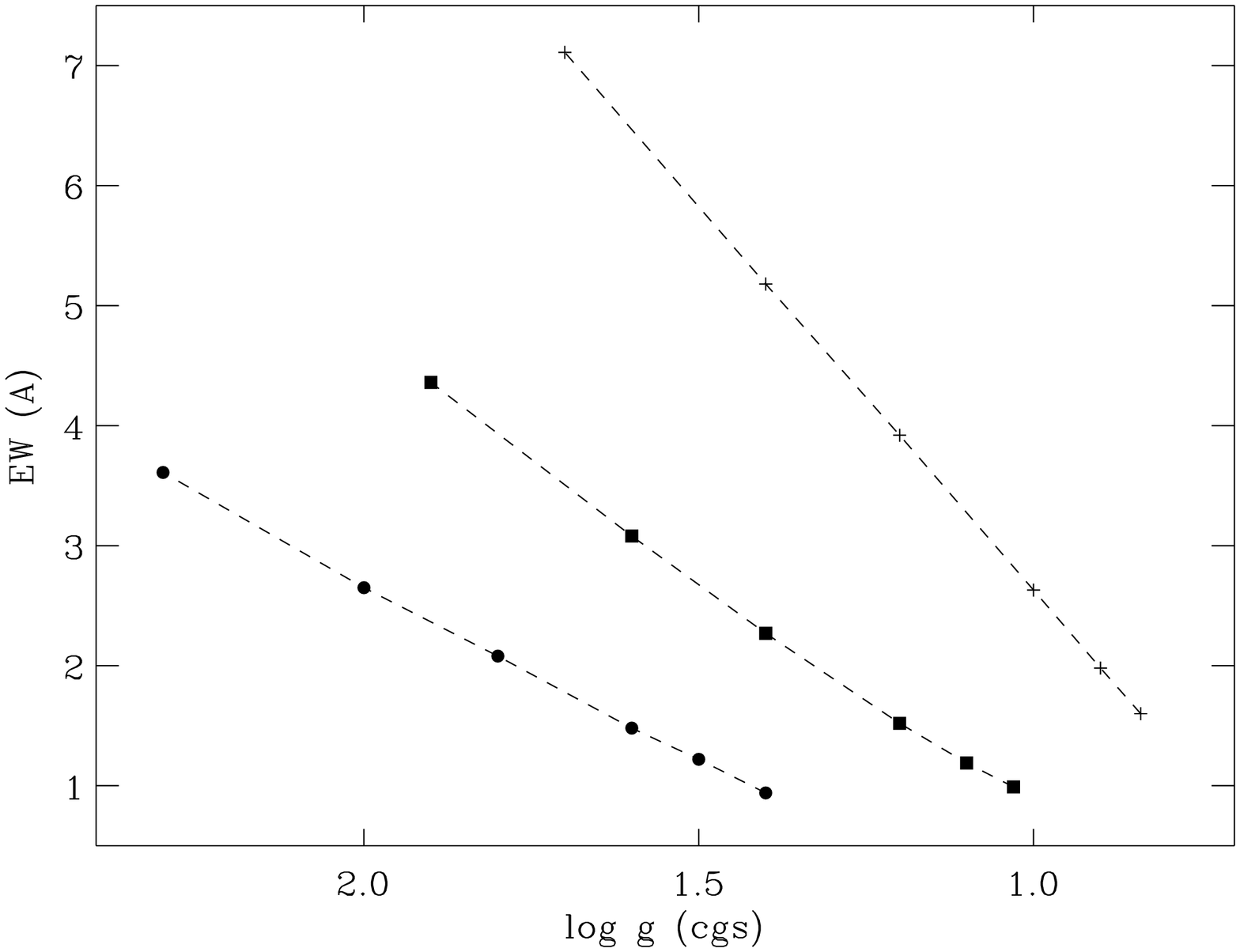}\\[-4mm]
\includegraphics[width=.4\textwidth,angle=90]{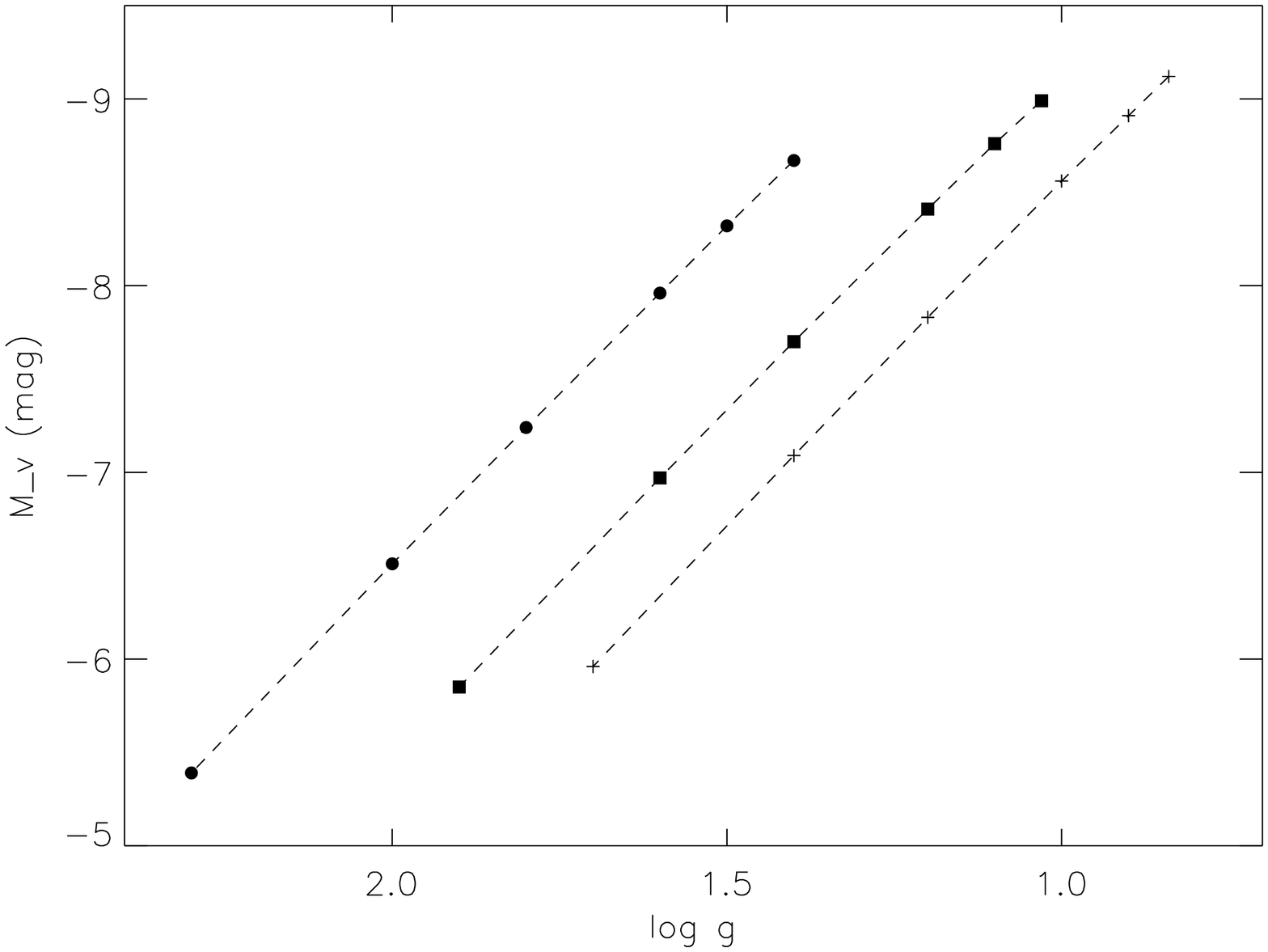}
\end{center}
\caption[]{
Three steps to calculate the theoretical correlation between absolute
magnitude $M_{V}$ and H$_{\gamma}$ equivalent width $EW$ for three different
values of $T_\mathrm{eff}$\,$=$\,8350\,K ({\it crosses}), 9500\,K ({\it
squares}), 12000\,K ({\it circles}). Top: $M_{V}$ vs. $EW$. Middle: 
$EW$ vs. $\log g$ as obtained from model atmospheres. 
Bottom: $M_{V}$ vs. $\log g$ as obtained from stellar evolution 
}
\label{ewmv}
\end{figure}

We illustrate the situation by using the stellar evolution models from
Fig.~\ref{hrd}. We select three values of $T_\mathrm{eff}$\,$=$\,12000, 9500 
and 8350\,K,
corresponding to spectral types B8, A0 and A4, respectively (see
\cite{bres03}, this volume), and calculate absolute visual magnitudes and
gravities for each track at each of the three $T_\mathrm{eff}$ values.
Figure~\ref{ewmv} displays the corresponding correlation of $M_{V}$ with $\log g$
for each temperature. While the correlations are nicely parallel, the
temperature dependence as a result of stellar evolution and bolometric
correction is obvious.

We can now calculate atmospheric models for each pair of $T_\mathrm{eff}$
and $\log g$
and plot calculated H$_{\gamma}$ equivalent widths $EW$ as function of gravity
and then, finally, produce a diagram of absolute magnitude versus 
H$_{\gamma}$ equivalent width. This is also done in Fig.~\ref{ewmv}. Since
the strengths of the Balmer lines do not only depend on gravity but also
on temperature, the final correlation
\begin{equation}
M_V = f(EW(\hgamma)\;,\bf{T_{eff}})
\end{equation}
depends very strongly on $T_\mathrm{eff}$. The equivalent widths are much
stronger at lower $T_\mathrm{eff}$ and, as an additional complication, the slope of 
the correlation is strongly temperature dependent. While this result is, at
least qualitatively, confirmed by observation \cite{bres03}, it is clear
that simple empirical magnitude -- equivalent width relations will have to 
suffer quite some intrinsic scatter unless it is restricted to accurate
spectral sub-types. We, therefore, suggest a method, which overcomes the
problem of the strong temperature dependence.

Before we do this important step, we draw attention to an important detail in
Fig.~\ref{ewmv}. The plot equivalent width versus $\log g$ reveals that gravities
can be determined with high precision for each effective temperature. An error
of $\sim$10~percent in $EW$ transforms to an error of 0.05 dex in $\log g$. Knowing
that we will have many higher Balmer lines in the blue spectra of supergiants
we feel confident that we can determine gravities with such accuracies, even
if the spectroscopic resolution is only moderate.

We now turn to the derivation of the \emph{Flux-Weighted Gravity -- Luminosity
Relationship} (FGLR), which was introduced very recently by~\cite{kud2003}. 
When discussing Fig.~\ref{hrd} in Sect.~1 we noted
that massive stars evolve through the domain of blue supergiants with
constant luminosity and constant mass. This has a very simple, but very
important consequence for the relationship of gravity and effective
temperature along each evolutionary track. From
\begin{equation}
L \propto R^{2}T^{4}_\mathrm{eff} = \mathrm{const.} ; M = \mathrm{const.}
\end{equation}
follows immediately that
\begin{equation}
M \propto g\;R^{2} \propto L\;(g/T^{4}_\mathrm{eff}) = \mathrm{const.}
\end{equation}

This means that each object of a certain initial mass on the ZAMS
has its specific value of the \emph{``flux-weighted gravity''
g/T$^{4}_\mathrm{eff}$} during the blue supergiant stage. 
This value is 
determined by the relationship between stellar mass and luminosity, which to
a good approximation is a power law
\begin{equation}
L \propto M^{x}\;.
\end{equation}

Inspection of evolutionary calculations with mass-loss, cf. \cite{meynet94}
and \cite{meynet00}, shows that $x=3$ is a good value in the range of
luminosities considered, although $x$ changes towards higher masses.
With the mass -- luminosity power law we then obtain
\begin{equation}
L^{1-x} \propto (g/T^{4}_\mathrm{eff})^{x}\;,
\end{equation}
or with the definition of bolometric magnitude 
$M_\mathrm{bol}$\,$\propto$\,$-2.5\log L$
\begin{equation}
-M_\mathrm{bol} = a\log(g/T_{\!\mbox{\scriptsize eff}}^4) + b\;.
\end{equation}

\begin{figure}[t]
\begin{center}
\includegraphics[width=.52\textwidth,angle=90]{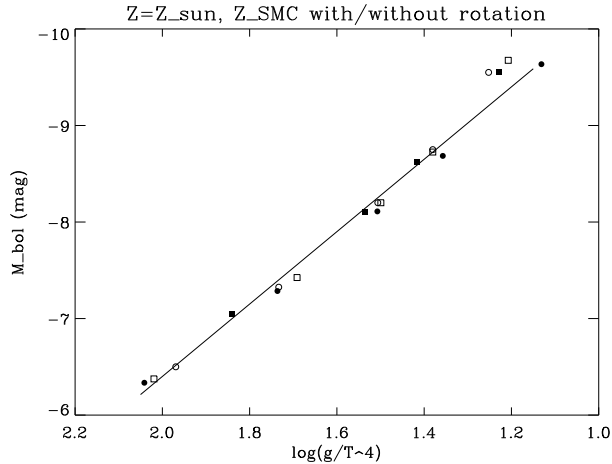}
\end{center}
\caption[]{
The FGLR of stellar evolution models from \cite{meynet94} and~\cite{meynet00}.
Circles correspond to models with rotation, squares represent models without
the effects of rotation. Solid symbols refer to galactic metallicity and
open symbols represent SMC metallicity. The solid curve corresponds to 
$a=-3.75$ in the FGLR
}
\label{fglrev}
\end{figure}

This is the FGLR of blue supergiants. Note that the proportionality constant
$a$ is given by the exponent of the mass -- luminosity power law through
\begin{equation}
a = 2.5 x/(1-x)\;.
\end{equation}
and $a=-3.75$ for $x=3$. Mass-loss will depend on metallicity and
therefore affect the mass -- luminosity relation. In addition, stellar rotation
through enhanced turbulent mixing might be important for
this relation. In order to investigate these effects we have used the models of
\cite{meynet94} and~\cite{meynet00} to construct the stellar evolution FGLR, 
which is displayed in Fig.~\ref{fglrev}. The result is very encouraging. All
different models with or without rotation and with significantly different
metallicity form a well defined very narrow FGLR.

\begin{figure}[t]
\begin{center}
\includegraphics[width=.98\textwidth]{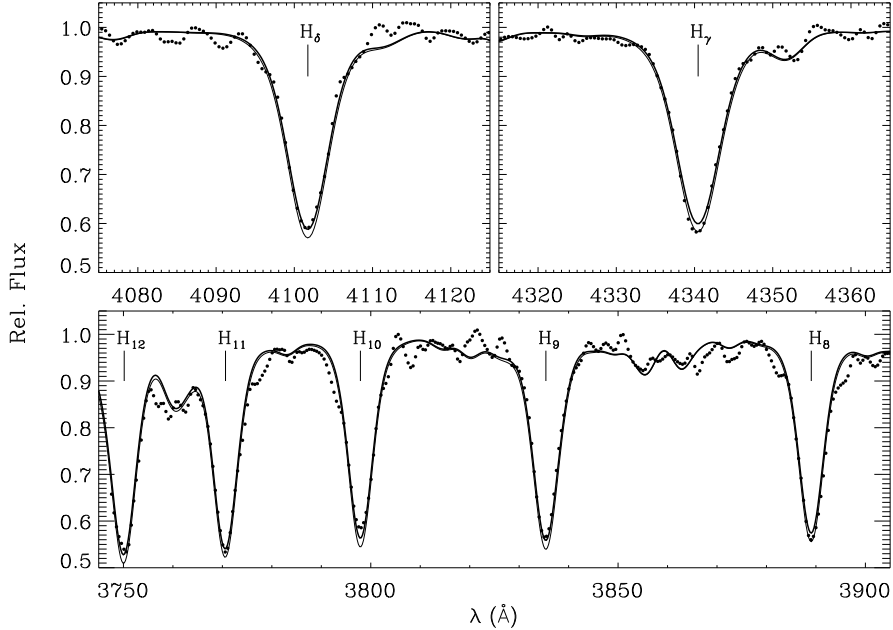}
\end{center}
\caption[]{
Fit of the higher Balmer lines of an A-supergiant in the Sculptor galaxy 
NGC\,300 using two atmospheric models with $T_\mathrm{eff}=9500$\,K and 
$\log g=1.60$ ({\it thick line}) and 1.65 ({\it thin line}), respectively. 
The data were taken with
FORS at the VLT. For further discussion see \cite{kud2003}
}
\label{balmfit}
\end{figure}

With this nice confirmation of our basic concept we discuss the possible
observational scatter arising from uncertainties in the determination of 
$T_\mathrm{eff}$ and $\log g$. As discussed in Sect.~2 and as also demonstrated by
\cite{kud2003}, effective temperature and gravity can be determined within
4 percent and 0.05\,dex, respectively. Treating these errors as independent
we derive an expected one sigma scatter $\Delta M_\mathrm{bol}=0.3$\,mag for the
FGLR per individual object, which is again very encouraging and suggests that the method, after
careful observational calibration (see \cite{kud2003} and \cite{bres03}, this
volume), might become a powerful distance indicator.

\begin{figure}[t]
\begin{center}
\includegraphics[width=.85\textwidth]{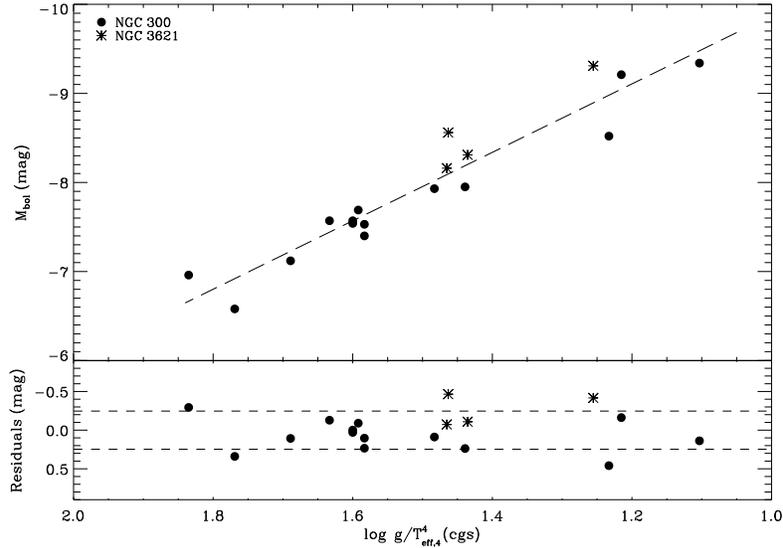}
\end{center}
\caption[]{
The FGLR of B8 to A4 supergiants in NGC\,300 and NGC\,3621; from 
\cite{kud2003}
}
\label{fglr1}
\end{figure}

Figure~\ref{balmfit} demonstrates how precisely the Balmer lines can be
fitted to yield a very accurate $\log g$ and Fig.~\ref{fglr1} shows the first
verification of the existence of a very tight FGLR for spiral galaxies
beyond the Local Group. Further results are shown in \cite{kud2003} and
\cite{bres03}.

\section{Conclusions}

We conclude that blue supergiants provide a great potential as excellent
extragalactic distance indicators. The quantitative analysis of their spectra
-- even at only moderate resolution -- allows the determination of stellar 
parameters, stellar wind properties and chemical composition with remarkable
precision. In addition, since the spectral analysis yields intrinsic energy
distributions over the whole spectrum from the UV to the IR, multi-colour
photometry can be used to determine reddening, extinction laws and extinction.
This is a great advantage over classical distance indicators, for which only
limited photometric information is available, when observed outside the Local
Group. Spectroscopy also allows to deal with the effects of crowding and
multiplicity, as blue supergiants, due to their enormous brightness,
are less affected by such problems than for instance Cepheids, which are
fainter.

Two tight relationships exist, the WLR and the FGLR, which can be used to
derive accurate absolute magnitudes from the spectrum with an accuracy of 
0.3\,mag per individual object. Applying the methods on objects brighter than
$M_{V}=-8$\,mag and using multi-object spectrographs at 8 to 10m-class
telescopes, which allow for quantitative spectroscopy down to
$m_{V}=22$\,mag, we estimate that with 20 objects per galaxy we will be
able to determine distances out to distance moduli of $m-M \sim30$\,mag
with an accuracy of 0.1\,mag. We note that these distances will not be affected
by uncertainties in extinction and metallicity, because we will be able to
derive the corresponding quantities from the spectrum.

\section{Acknowledgements}
It is a pleasure to thank the organizers of the workshop for a very
stimulating meeting. We gratefully acknowledge the hospitality and support
of the local team. We also wish to thank our colleagues Fabio Bresolin
and Wolfgang Gieren for their significant contributions to this work.

%\clearpage

%INDEX%%%%%%%%%%%%%%%%%%%%%%%%%%%%%%%%%%%%%%%%%%%%%%%%%%%%%%%%%%%%%%%
% Please check with the editor of your book whether he plans to
% include a "mutual" subject index - if so, please code your entries
% in the standard syntax. For your own purposes you may print your
% "personal" index by using the following commands:
%
%\clearpage
%\addcontentsline{toc}{section}{Index}
%\flushbottom
%\printindex
%%%%%%%%%%%%%%%%%%%%%%%%%%%%%%%%%%%%%%%%%%%%%%%%%%%%%%%%%%%%%%%%%%%%%

%\clearpage
%\addcontentsline{toc}{section}{Index}
%\flushbottom
%\printindex

\end{document}